# Dynamics under radiation damping from ab-initio formalism -nonlocal restraint on acceleration runaway in electrodynamics

D. Das*

(dasd1951@gmail.com)
Bhabha Atomic Research Centre
Trombay, Mumbai 400 085

Abstract

Classically encountered problem of acceleration runaway in radiative motion of a charged-particle is addressed by considering the criterion of minimum radiation loss under optimum displacement in variationally referred dynamic course in between a pair of arbitrarily selected space-like surfaces having time-like separation. Using the criterion, analysis made by involving a family of infinitesimally differed paths reveals that for attaining stationary action of the dynamic course it is necessary to involve orchestrated displacement of the path family. Dynamics that generally follows from such analysis involves nonlocally mediated interaction besides the conventional ones. Amidst external field perturbation it is shown that ubiquitous nonlocal mediation incessantly attempts optimizing displacement in accomplishing minimum radiation loss. The mediation involves virtual exchange of quantized energy and momentum of the interacting field and particle, which in conformity with quantum mechanical prescription is describable by unitary evolution. Under increasingly high perturbation, the mediation can boost its defensive effort but up to a critical limit with stress release predominantly as Larmor radiation relaxation. Observable acceleration and jerk, and relaxations thereof in the nonstationary evolution are indeed markers of the defense failure. Recoil from the Larmor relaxation impedes kinetic energy growth and also imparts shear stress to the spherically polarized vacuum field of point charge; the transverse stress release having similitude of Hawking-Unruh radiation. Defending nonlocal field restrains acceleration runaway and loss of causality. The classically based ab-initio analysis on nonlocal realisms further proves its versatility in revealing nonlocally mediated geodesics structure of quantized 4-space with useful revelations.

Key words: Lorentz-Abraham-Dirac equation, Nonlocal influence, Coherence and decoherence, Hawking radiation.

*Retired scientist from Bhabha Atomic Research Centre, Trombay, Mumbai.

**Contents**

## 1. Introduction

Radiation emission from an accelerated charge and its recoil effect in the motion has drawn attentions of many investigators [1-8] while addressing to the notorious acceleration runaway problem encountered in the dynamical analysis. Radiation related force classically derived from detailed analysis of self field loss of the accelerated charge depends on jerk ($\ddot{\bar{v}}$) involved in the motion. In external electromagnetic field ($\bar{F}_{\text{external}}$), the derived dynamics, $m_0\dot{\bar{v}} = \bar{F}_{\text{external}} + (2q^2/3c^3)\ddot{\bar{v}}$, expresses the growth of inertial momentum ($m_0\dot{\bar{v}}$) in the presence of radiation force, $(2q^2/3c^3)\ddot{\bar{v}}$; $q$ and $m_0$ are charge and mass of the particle, and $c$ is signal speed. $\bar{F}_{\text{external}} = [q\bar{E} + (q/c)(\bar{v}\times\bar{H})]$, $\bar{E}$ and $\bar{H}$ being the electric and magnetic field components of the external field. According to the dynamics, when external field is withdrawn abruptly, the inertial momentum continues to grow up due to the jerk dependent force $(2q^2/3c^3)\ddot{\bar{v}}$. In the post withdrawal period of external field, the dynamics predicts exponential growth of acceleration with the time constant of $(2q^2/3m_0c^3) \sim 10^{-24}\text{s}$. Exponential growth of acceleration in the field free domain hints at some shortcoming in the considerations that are involved in deriving the radiation force. Nature of shortcoming can be understood while analyzing the effects of direct and transverse components of jerk force. Jerk influencing upon radiative motion involves the two mutually orthogonal force components: $(2q^2/3c^3)\ddot{\bar{v}} \equiv (2q^2/3c^3)[\ddot{v}\bar{e} + \dot{v}\dot{\bar{e}}]$, where, $\ddot{\bar{v}} \equiv d(\dot{v}\bar{e})/dt$; $\bar{e}$ and $\dot{\bar{e}}$ respectively being the unit acceleration vector ($\bar{e} = \dot{\bar{v}}/\dot{v}$) and its angular sweeping (torsion) rate ($\dot{\bar{e}}$). In general, time evolution of the normal can be written as $\dot{\bar{e}} = \sum_{k,\omega_k} \bar{e}_k(-i\omega_k)\exp(-i\omega_k t)$, $\sum_k \bar{e}_k \bullet \bar{e}_k^* = 1$, ($\bar{e}_k \bullet \bar{e}_{k'}^* = \delta_{kk'}$, $\delta_{kk'}$ is Kronecker delta), where $\omega_k$'s are internal frequencies of the angular evolution, ($\bar{k}$ being wave vector). Angular velocity $\dot{\bar{e}}$ is ascribed to the evolution of inherently present spherically polarized vacuum field of the point charge with dynamically relevant radius of $r_0 = 2q^2/3m_0c^2$ and evolution period of $4t_0$, ($t_0 = r_0/c$); $\omega_0 = 2\pi/4t_0$ representing averaged evolution frequency. Electrodynamics with constant acceleration also allows finite evolution of the normal.

The direct and transverse components of the jerk force should be dynamically balanced by remaining force terms of electrodynamics. Heuristically, one expects that the jerk force component, $(2q^2/3c^3)\dot{v}\dot{\bar{e}}$, implies torsional stress on angularly evolved normal of the polarized charge sphere, and that it is counterbalanced by recoil component of the transverse electromagnetic field from the emitting Larmor radiation of accelerated charge. The spherically polarized vacuum field of moving charge experiences the torque stress of $\bar{F} \equiv (2q^2/3c^3)\dot{v}\dot{\bar{e}} = \bar{\Gamma}\dot{e}/c$, where, $\bar{\Gamma} \equiv [r_o\bar{n}_r \times m_0\dot{\bar{v}}] = r_o m_0\dot{v}\bar{\varepsilon}$, ($\bar{\varepsilon} \equiv \bar{\dot{e}}/\dot{e} = \bar{n}_r \times \bar{e}$ $\bar{n}_r$ being unit radial vector). Heuristic argument thus points out the necessity of explicit involvement of radiation recoil force in the dynamic representation. The explicit form would then help analyze possible dynamic solution of jerk effect under abrupt withdrawal of external field, in which case, radiation emitting with the time constant of about $10^{-24}$s continues imparting reaction to the direct as well as transverse components of jerk force. In the conventional electrodynamics, recoil effect of the transverse electromagnetic field on jerk force components is not explicitly analyzed. It is a shortcoming indeed. As will be elaborated later in the text, the radiation recoil is an essential feature in the nonstationary motion of electrodynamics that involves radiation relaxation in the neighborhood of field-particle interaction zone of average radius $r_0$. Classical analysis does recognize the regular (Larmor) radiaton loss and grossly considered the energy-momentum loss of accelerated charge at the radiation boundary, but the analysis remained oblivious of nonlocally mediated governance of field-particle interaction involved within the said boundary. In the presence of indicated shortcoming, classical analysis could not also realize the occurrence of additional power loss in radiation relaxation from the torsional stress in nonstationary evolution that occurs under critical failure of nonlocally defended coherence in field-particle interaction. The additional power loss becomes apparent as one considers the impact momentum due to radiation recoil on charged particle of mass $m_0$.

Recoil-impact within the characteristic time $t_0$ results in transverse velocity component to be assigned as, $\bar{v}_{trans} \propto -\dot{\bar{\Pi}} t_0 / m_0 \equiv -(2q^2\dot{v}^2\bar{n}_{rad}/3c^4)t_0/m_0$, where, $\dot{\bar{\Pi}}$ is the rate of momentum loss in Larmor radiation ($\bar{n}_{rad}$ being unit radiation vector). The proportionality indicates that a part of the recoil impact is effective in counteracting the transverse jerk component, (remaining part counteracting on direct component of jerk force). It is interesting to note that the implied power loss proportional to $(-\dot{\bar{\Pi}}) \bullet \bar{v}_{trans} \equiv \dot{\bar{\Pi}}^2 t_0 / m_0$, is having $\dot{v}^4$ dependence, which is different in nature as compared to the $\dot{v}^2$ dependence of Larmor radiation loss. The power loss in transverse mode is functionally distinct from those involved in the direct displacement based power balance considered in the classical dynamics as $\dot{k} = \bar{F}_{external} \bullet \bar{v} + (2q^2/3c^3)\ddot{\bar{v}} \bullet \bar{v}$, (k being the kinetic energy), where, $\ddot{\bar{v}} \bullet \bar{v} \equiv [\ddot{v}\bar{e} \bullet \bar{v} + \dot{v}\dot{\bar{e}} \bullet \bar{v}] = \ddot{k}/m_0 - \dot{v}^2$. Power term, $(2q^2/3c^3)\ddot{v}\bar{e} \bullet \bar{v}$, excludes torsion based angular displacement because of the equality, $\ddot{v}\bar{e} \bullet \bar{v} = \ddot{k}/m_0 - (\dot{v}^2 + \dot{v}\dot{\bar{e}} \bullet \bar{v})$. Transverse displacement signifying dissipative (disorderly) power loss cannot figure in the direct displacement related power, which accounts for the impeded kinetic energy growth ($\dot{k}$-$t_0\ddot{k}$) and the orderly Larmor radiation loss as ($\dot{k}$-$\tau_0\ddot{k}$) + $(2q^2/3c^3)\dot{v}^2$ = $\bar{F}_{external} \bullet \bar{v}$. It is imperative that the external field to sustain accelerated motion of charge in the presence of orderly, and also, the indicated dissipative relaxations, bears all the relaxation losses.

The above stated unexplored relaxation loss arising from radiation reaction on transverse component of the jerk force needs further attention as it might have underlying connection with Hawking-Unruh type radiation emission [9,10] density that has also biquadratic dependence on acceleration. It is again noteworthy that in the case of uniformly accelerated charged particle although jerk has null expectation value ($\langle \ddot{\mathrm{v}} \rangle = 0$), the transverse force component, $(2\mathrm{q}^2/3c^3)\langle \dot{\mathrm{v}}\bar{\dot{\mathrm{e}}} \rangle \equiv t_0 \langle m_0 \dot{\mathrm{v}}\bar{\dot{\mathrm{e}}} \rangle$, does have finite magnitude in the nonstationary situation of steady growth of kinetic energy. In this particular case, while nonzero observable acceleration ($\langle \dot{\mathrm{v}} \rangle \neq 0$) accounts for the regular (Larmor) radiation emission, the nonzero $\langle \dot{\mathrm{v}}\bar{\dot{\mathrm{e}}} \rangle$ points to disordered (dissipative) radiation emission. Disordered emission is generally negligible as compared to the regular one. This aspect will be detailed later in the text.

Introspection into the classical formulation indeed reveals that the formulation recognizes the radiation force only from the viewpoint of energy-momentum conservation in the presence of momentum loss in Larmor radiation emission. But it does not look into the need of having a measure of optimized loss of radiation energy and momentum from the accelerated charge in completing the dynamic passage referred in the variation analysis for stationary action. In the analysis, although the infinitesimally differed paths $\{x_p^\alpha\}$ of stationary action are considered within a pair of space-like boundaries having time-like separation, the instant displacements of respective paths need not remain optimum. The classical world path arrived thereby need not endorse minimum power loss in the stated dynamic course. In a dynamic course with an overall displacement of $\Delta\tau \equiv \tau_2 - \tau_1$, the radiation power loss given by $\int_{\tau_1}^{\tau_2} (dR/d\tau)d\tau$, $dR/d\tau = (2q^2/3)\dot{v}^2$, (that is, $dR/dt = (2q^2/3c^3)\dot{\mathrm{v}}^2$) would remain minimum when one considers that path evolution occurs with shortest segment length to be expressed as $\delta\int_{\tau_1}^{\tau_2} (dx_p^\alpha dx_p^\beta g_{\alpha\beta})^{1/2} = 0$, where $v^\alpha \equiv dx^\alpha/d\tau$ is 4-velocity and $\dot{v}^\alpha$ is 4-acceleration ($\dot{v}^2 \equiv \dot{v}^\alpha \dot{v}_\alpha$). In instant commoving inertial frame, $\dot{v}^2 = -\dot{\mathrm{v}}^2/c^4$. As the evolution meets the stated displacement criterion, self field loss at the radiation boundary will remain minimized. As will be seen shortly, the criterion would implicate involvement of nonlocally mediated field-particle coherent interaction leading to optimum relaxation as radiation. The mediation feature is anyway absent in the classical analysis, because of which the causally interconnected relaxation events at radiation boundary remain subjective to the displacement history. The incompleteness is reflected by the well marked presence of unwanted acausal consequence like acceleration runaway under withdrawal of external power source abruptly. The nonlocal mitigation remaining unaccounted in classical analysis, the power loss ultimately borne by the external field is left open without a measure. Classically described radiation related force being solely based on energy-momentum conservation it falls short of the stated measure. Nonlocal mediating feature that optimizes the relaxation loss in the accelerated dynamics can be explored by using the classical Lagrangian approach of variational formalism involving stationary action in a dynamic passage. Next section shows that by introducing the optimization criterion within the classical approach, local and nonlocal features get abridged with revelation of nonlocally mediated dynamic evolution characteristics seamlessly. The abridgement does not called for axiomatic introduction of operator representations of canonical quantities in dynamics. The formalism presented in this study includes detailed mediation characteristics of nonlocal field in coherence and in critical loss of coherence.

## 2. Classically explored nonlocal interplay in electrodynamics

-"local and nonlocal realisms in electrodynamics"

In the variationally defined dynamic course of stationary action in between a pair of arbitrarily selected space-like surfaces with their time-like separation, radiation loss (or, gain) can be generally optimized when the course is endowed with features of canonically orchestrated minimum displacement of linearly combined infinitesimally differed paths, $\{x_p^\alpha(\tau)\}$. As shown below, the orchestrated displacement rule that helps in attaining minimally averaged passage length (displacement) is necessarily mediated through nonlocal means of well defined delocalized evolution. Nonlocal field ubiquitously manifests to mediate the dynamic properties coherently in field-particle interaction. Ubiquitous involvement of nonlocal governance is later shown to be deeply connected with the omnipresent vacuum field (see in appendix). The nonlocally governed delocalized dynamic passage accomplishes optimum power loss whenever radiation relaxation occurs during displacement. Infinitesimal displacement of the linearly combined paths defining the passage will be expressed by $d\tau \equiv [dx^\alpha dx^\beta g_{\alpha\beta}]^{1/2}$ $= \left( \sum_p c_p^2 dx_p^\alpha dx_p^\beta \right)^{1/2}$, where $dx^\alpha = \sum_p c_p dx_p^\alpha$ ; $g^{\alpha\beta}$ being metric of the dynamic 4-space with signature (1,-1,-1,-1). The weight factors $\{c_p\}$ of the linearly combined displacements of respective paths are ruled by coherent mediation characteristics of nonlocal field; $c_p c_{p'} = \delta_{pp'}$, $\delta_{pp'}$ being Kronecker delta, $\sum_p c_p^2 = 1$. The rule endorses stationary action in the course as $\int_{\tau_1}^{\tau_2}(\delta L)d\tau = 0$, where, $\delta L = \sum_p c_p^2 \delta L_p$ $L_p \equiv L_p(x_p^\alpha, v_p^\alpha)$ defining Lagrangian on path, $p$ of the family $\{x_p^\alpha(\tau)\}$; $\delta L_p = [(\partial L_p / \partial x_p^\mu)\delta x_p^\mu + (\partial L_p / \partial v_p^\mu)\delta v_p^\mu]$.

The involvement of nonlocal connectivity in the orchestrated evolution of dynamic course will be apparent from optimization of the linearly averaged segment length of the delocalized passage to be expressed as $\delta\int_{\tau_1}^{\tau_2}[\sum_p c_p^2 dx_p^\alpha dx_p^\beta g_{\alpha\beta}]^{1/2} = 0$. The criterion of shortest segment length that optimizes radiation loss/gain in the course implies the following characteristics:

$$v_\beta \delta x^\beta \Big|_{\tau_1}^{\tau_2} - \int_{\tau_1}^{\tau_2} \delta x^\beta d\tau[\dot{v}_\beta - v^\alpha v^\sigma \partial_\beta g_{\alpha\sigma} / 2] = 0 \qquad \text{(A1)},$$

where, $\dot{v}_\beta = \dot{v}^\alpha g_{\alpha\beta}$, $\dot{v}^\alpha(\tau)$ and $\delta x^\alpha$ respectively representing the sets $\dot{v}^\alpha \Leftrightarrow \{\dot{v}_p^\alpha\}$ and $\delta x^\alpha \Leftrightarrow \{\delta x_p^\alpha\}$ as $\dot{v}^\alpha(\tau) \equiv \sum_p c_p \dot{v}_p^\alpha(\tau)$ and $\delta x^\alpha \equiv \sum_{p'} c_{p'} \delta x_{p'}^\alpha$. One finds that the integrated term in eq.(A1) is a null by definition of the time-like boundaries at $\tau_1$ and $\tau_2$. This is so as the scalars $v_\beta \delta x^\beta$ at the respective limits are decided by variations of proper time ($\tau$) $(\delta x^0)_{\text{proper}}$, which are null by definition of the variation boundaries. Now, if the variations $\delta x^\beta$ are declared arbitrary, then eq.(A1) leads to equation describing geodesic paths as $\dot{v}_\sigma = v^\alpha v^\beta \partial_\sigma g_{\alpha\beta} / 2$. This classical result is in no way connected with the objective of analyzing the dynamic course with optimum displacement of a charged particle in external electromagnetic field. The set objective can, however, be achieved when one considers that eq.(A1) generally leads to 4-orthogonal connection involving argument of the integral that concerns the nonlocally introduced variations $\delta x^\alpha \Leftrightarrow \{\delta x_p^\alpha\}$. The 4-orthogonal connection stipulates the incessant mediation of ubiquitous nonlocal field in attaining the optimized displacement, $dx^\alpha = \sum_p c_p dx_p^\alpha$, using coherently linear combination of the paths $\{x_p^\alpha(\tau)\}$; nonlocally entangled paths forming the combination. The connection, $(\dot{v}_\sigma - v^\alpha v^\beta \partial_\sigma g_{\alpha\beta} / 2)\delta x^\sigma = 0$, or, $(Dv_\sigma / D\tau)\delta x^\sigma = 0$, is generally valid irrespective of the arbitrarily selected variation boundaries. The 4-orthogonal connection can be rewritten by replacing the set of variational elements $\delta x^\alpha \Leftrightarrow \{\delta x_p^\alpha\}$ with space-like unit 4-vectors, $i^\alpha \Leftrightarrow \{i_p^\alpha\}$, $i^\alpha \equiv \sum_{p'} c_{p'} i_{p'}^\alpha$, $(i^\alpha i_\alpha = -1)$. In flat space ($\partial_\sigma g_{\alpha\beta} = 0$), the orthogonal connection assumes the simpler form, $\dot{v}_\alpha i^\alpha = 0$. The nonlocal entanglement anyway continues to remain valid for the particular case where external field (for example, electromagnetic, nuclear, etc.) asymptotically disappears and the particle moves freely in the affine space.

Now, the nonlocal quantities $i^\alpha \Leftrightarrow \{i_p^\alpha\}$ being independent of displacement the 4-orthogonal connections (namely, $\dot{v}_\alpha i^\alpha = 0$ in flat space and $(Dv_\sigma / D\tau) i^\sigma = 0$ in curved space) generally suggest that the product $v_\alpha i^\alpha$ evolves irrespective of the displacement. Recalling that the family of entangled geodesics is defined in between a pair of arbitrarily selected time-like variation boundaries where the scalar product $v_\alpha \delta x^\alpha$ (precursor of $v_\alpha i^\alpha$) at either boundary is a null, the displacement independent evolution corroborates to the 4-orthogonal characteristics $v_\alpha i^\alpha = 0$ of the nonlocally mediated geodesics. By replacing $i^\alpha \equiv [\mathrm{i}^0, \bar{\mathrm{i}}] = \mathrm{i}^0[1, \bar{\mathrm{i}} / \mathrm{i}^0]$ with $w^\beta \equiv p_0[1, \bar{\mathrm{w}} / c]$, ($p_0$ proportional to $\mathrm{i}^0$), the arrived characteristics is recognizable as 4-orthogonal connection of displacement components with (nonlocal) velocity like components $w^\beta$, where, $\bar{\mathrm{w}} / c = \bar{\mathrm{i}}/\mathrm{i}^0 = \delta \bar{x} / \delta x^0$, $\bar{x} \equiv \sum_p c_p \bar{x}_p$, $x^0 \equiv \sum_p c_p x_p^0$ ,. Rewriting the characteristics as $v_\beta w^\beta = 0$, or, as $\bar{\mathrm{v}} \bullet \bar{\mathrm{w}} = c^2$ one finds that use of relativistic speed limit for the displacement rate gives the possible range of the nonlocal velocity ($\mathrm{w}$), or, phase velocity values, as $c \le \mathrm{w} \le \infty$ . Now, in the spectral space, $k^\beta \equiv [\omega / c, \bar{\mathrm{k}}]$, ($\bar{\mathrm{k}} = \mathrm{k}\bar{\mathrm{n}}_\mathrm{k}$ , $\mathrm{n}_\mathrm{k}^2 = 1$), the 4-orthogonal characteristics can be written as $v_k^\beta w_{k,\beta} = 0$, where, $w_k^\beta \equiv p_{0,\mathrm{k}}[1, (\omega/c\mathrm{k})\bar{\mathrm{n}}_\mathrm{k}]$. By comparing this result with the perennial property of phase velocity as $k^\beta w_{\mathrm{k},\beta} \equiv 0$, one infers that the nonlocally mediated entangled geodesics have their evolution property of $v_k^\beta \propto k^\beta$. Using the identical value of two summations, namely, $\sum_k c_k^2 v_k^\beta v_{k,\beta} = 1 = \sum_k c_k^2$ , one finds that arrived evolution property corroborates to $\sum_k c_k^2 [k^\beta k_\beta - (1/\lambda\!\!\!^-)^2] = 0$, where $1/\lambda\!\!\!^-$ serves as constant of the indicated proportionality. Thus, the relation, $k^\beta k_\beta \equiv \omega^2 / c^2 - \mathrm{k}^2 = 1/\lambda\!\!\!^-{}^2$, represents spectral dispersion characteristics of evolution of the nonlocally mediated geodesics. The finding reveals that the affine space is describable with the geodesics network made of the quantized dimension $\lambda\!\!\!^-$ and with quantized energy and energy density of $2\pi c\hbar / \lambda\!\!\!^-$ and $2\pi c\hbar / \lambda\!\!\!^-{}^4$ respectively. The quantized space features will be briefly discussed in Appendix 1.

As for understanding the dynamic course of a charged particle in external field, the result of displacement optimization suggests that the evolution of entangled geodesics' family stands for the particular case of dynamic evolution of free particle because of the similar dispersion characteristics. (Free particle evolution characteristics will be elaborated later as special case of the dynamic evolution in the absence of external field). In presence of external field the evolution deviates from that of entangled geodesics. Now, recalling that stationary action is expressible as $\int_{\tau_1}^{\tau_2} \{\sum_p c_p^2 \delta L_p\} d\tau = 0$, one writes

$$\int_{\tau_1}^{\tau_2} \left[ \sum_p c_p^2 (\partial L_p / \partial x_p^\mu + d\pi_{\mu,p} / d\tau) \delta x_p^\mu \right] d\tau - \left[ \sum_p c_p^2 (\pi_{\mu,p} \delta x_p^\mu) \right] \Bigg|_{\tau_1}^{\tau_2} = 0,$$

where, $\pi_{\alpha,p} \equiv -(\partial L_p / \partial v_p^\alpha)$, is the canonical 4-momentum on path $p$; $\pi^\alpha = \sum_p c_p \pi_p^\alpha$. Now, the integrated term of left hand side in the above equation will be absent when one stipulates the supplementary condition, $\sum_p c_p^2 (\pi_{\mu,p} \delta x_p^\mu)_{\text{at } \tau_1} = \sum_p c_p^2 (\pi_{\mu,p} \delta x_p^\mu)_{\text{at } \tau_2}$. Under this stipulation the resultant equation shows that stationary action principle endorses dynamic evolution of the coherently displacing paths $\{x_p^\alpha(\tau)\}$ irrespective of variation boundaries as

$$(\partial L / \partial x^\mu + d\pi_\mu / d\tau) i^\mu = 0 \qquad \text{(A2).}$$

The supplementary condition as stipulated for obtaining eq.(A2) is considered now in rewritten form as

$$(\pi_\alpha \delta x^\alpha)_{\tau_1} = (\pi_\alpha \delta x^\alpha)_{\tau_2} = C \text{ (say)} \qquad \text{(A2a).}$$

The equality (A2a) if taken to be valid irrespective of the arbitrarily selected variation boundaries, the parameter $C$ can be evaluated by considering a special initial boundary at $\tau_1$, where the external field asymptotically disappears and the particle assumes free evolution corroborating to $\pi^\alpha = m_0 c v^\alpha$. For free motion, noting that the displacement $v_\alpha$ occurs along geodesics of 4-space involved in the dynamic description, and also recalling that the geodesics evolution as $v_\alpha i^\alpha = 0$, one writes the equality, $(\pi_\alpha \delta x^\alpha)_{\tau_1} \equiv m_0 c (v_\alpha i^\alpha)_{\tau_1} = 0$. In absence of external field at the selected initial boundary, the dynamic evolution characteristics corroborates to that of nonlocally entangled geodesics. Noting that such initial boundary selection is generally a feasibility as dynamic history, the parameter $C$ of the supplementary condition (A2a) is a null and it implies that dynamic evolution corroborates to the 4-orthogonality for all $\tau = \tau_2$.

$$\pi_\alpha i^\alpha = \pi_\alpha w^\alpha = 0 \qquad \text{(A2b).}$$

Eq.(A2b) generally applies to the coherently mediated evolution of path family $\{x_p^\alpha(\tau)\}$ irrespective of the arbitrarily selected boundary. As will be elaborated in section 3, the mediating feature, $i^\alpha$ (or $w^\alpha$) endorses quantum nature of canonical energy-momentum involved in field-particle interaction. For free particle case, the characteristics, $v_\beta w^\beta = 0$, or, $\bar{\mathrm{v}}\bullet\bar{\mathrm{w}} = c^2$ is reminder of the association of deBroeglie wave evolution in free motion.

In view of displacement independency of the mediator, Eq.(A2b) further leads to $\dot{\pi}_\alpha i^\alpha = \dot{\pi}_\alpha w^\alpha = 0$. Noting this, one realizes that the null value of the product $(\partial L/\partial x^\mu + d\pi_\mu / d\tau) i^\mu$ expressed by eq.(A2) will imply the orthogonal connection $(\partial L/\partial x^\alpha) i^\alpha = 0$ when the interacting field $-\partial L/\partial x^\alpha$ is not externally controlled and it evolves under the governance of nonlocal mediator. However, when the field is externally driven, the nonlocal governance on it is generally lost. The canonical 4-momentum expressed in (A2b) will evolve in a dynamic course where mitigating reaction of the governor on the external perturbation critically manifests against the externally driven dynamics which is expressible as $\dot{\pi}_\alpha + \partial L/\partial x^\alpha - f_\alpha = 0$. Added 4-force, $-f_\alpha$, is signature of mitigating governance. As will be seen later in the text that finite $-f_\alpha$ corroborates to radiation reaction 4-force in electrodynamics. Its manifestation is marker of nonstationary dynamic evolution if charge particle, which occurs under critically high externally perturbing electromagnetic field. Critical attainment of the nonlocal defense to the perturbation will be more clear once the radiation 4-force $f_\alpha$ is fully explored (refer subsection 2.1.2).

As for the functionality of $f^\alpha$ one considers the requirement of its space-like characteristics, $f^\alpha v_\alpha = 0$, which helps conserving rest mass of the charged particle in electrodynamics that follows from the equation, $\dot{\pi}_\alpha = -\partial L/\partial x^\alpha + f_\alpha$. Using the Lagrangian, $L = -m_0 c (v^\alpha v_\alpha)^{1/2} - (q/c) A^\alpha v_\alpha$, ($A^\alpha$ being components of 4-potential of external electromagnetic field), the derived electrodynamics having the form, $d(m_0 c v_\mu)/d\tau - (q/c) F_{\mu\nu} v^\nu - f_\mu = 0$, ($F_{\mu\nu} = \partial_\mu A_\nu - \partial_\nu A_\mu$), conforms to the required characteristics of unchanged proper mass as $f^\alpha v_\alpha = d(m_0 c)/d\tau = 0$. Now the conventionally considered radiation 4-force expressed by Lorentz-Abraham-Dirac term $(2q^2/3c)[\ddot{v}^\alpha + \dot{v}^2 v^\alpha]$ although endorses the space-like characteristics as $[\ddot{v}^\alpha + \dot{v}^2 v^\alpha] v_\alpha = 0$, it does not have any nonlocally governed term to optimize displacement in the presence of radiation relaxation and recoil effect thereof. Any attempt to include a nonlocally governed term as modification of the LAD 4-force must conform to the stated requirement of space-like characteristics. The possible nonlocally governed term to be associated with LAD 4-force is now explored by considering the optimum displacement criterion eq.(A1). The nonlocal feature is worked out in the following subsection.

## 2.1 **Radiation 4-force as noncally modified Lorentz-Abraham-Dirac term**

For inducting the nonlocal modification wit LAD 4-force, one recalls displacement optimization criterion expressed in eq.(A1). For flat space, eq.(A1) expresses 4-orthogonal connection, $\dot{v}_\alpha i^\alpha = 0$. There, one finds that the scalar product $v_\alpha i^\alpha$ evolves irrespective of displacement in presence of external field. The displacement independent evolution can be written generally in 4-orthogonal form as $v_\alpha i'^\alpha = 0$, by defining the space-like 4-vector, $i'^\alpha \equiv i^\alpha - v_\otimes v^\alpha$, where $v_\otimes \equiv v_\alpha i^\alpha$. Scalar product, $v_\otimes$, as seen already, is a null for the particular case of nonlocally mediated evolution of free particle. For general case, noting the 4-orthogonal connection, $v_\alpha i'^\alpha = 0$, one explores involvement of the nonlocally influenced 4-vector $i'^\alpha$ as added modification over the conventional LAD term to express the radiation 4-force as $f^\alpha = (2q^2/3c)[(\ddot{v}^\alpha + \dot{v}^2 v^\alpha) + \Theta i'^\alpha]$; $\Theta$ being a scalar to be normalized. The resultant 4-force anyway endorses the space-like characteristics, $f^\alpha v_\alpha = 0$. Furthermore, $f^\alpha$ also endorses its null 4-projection on $i'^\alpha$ because of the fact that no other terms in the electrodynamics expressed by $d(m_0 c v_\mu)/d\tau - (q/c)F_{\mu\nu}v^\nu - f_\mu = 0$ are influenced nonlocally. ($i^\alpha$ as well as $i'^\alpha$ is 4-orthogonal to 4-acceleration and bears null 4-projection on the locally defined 4-force term $(q/c)F_{\mu\nu}v^\nu$). One thus express this characteristic as

$$f_\alpha i'^\alpha = f_\alpha i^\alpha = 0 \qquad \text{(A2c).}$$

Now, substituting $f^\alpha$ as the modified LAD term in eq.(A2c), and then using the two equalities, namely, $\ddot{v}^\alpha i_\alpha \equiv d(\dot{v}^\alpha i_\alpha)/d\tau = 0$, and $i'^\alpha i_\alpha = i'^\alpha i'_\alpha = -(1+v_\otimes^2)$, one finds that the scalar $\Theta$ is obtainable from null value of the product, $[(\ddot{v}^\alpha + \dot{v}^2 v^\alpha) + \Theta i'^\alpha] i_\alpha \equiv \dot{v}^2 v_\otimes - \Theta(1+v_\otimes^2) = 0$. Thus, by using the displacement optimization criterion the conventionally arrived LAD 4-force gets modified into local-nonlocal abridged form as $f^\alpha = (2q^2/3c)[(\ddot{v}^\alpha + \dot{v}^2 v^\alpha) + \Theta i'^\alpha]$, where $\Theta = v_\otimes \dot{v}^2/(1+v_\otimes^2)$. In field free evolution (4-jerk and 4-acceleration are absent, and $v_\otimes$ is null) the modified form $f^\alpha$ is a null.

It is worthwhile to mention here that the above stated modification could have been obtained alternatively by considering that $f^\alpha$ is a linear combination of the three independent dynamically relevant features, namely, reduced 4-velocity $v'^\alpha \equiv v^\alpha + v_\otimes i^\alpha$, ($v'^\alpha i_\alpha = 0$; $i^\alpha i_\alpha = -1$), 4-acceleration and 4-jerk. That is, $f^\alpha = A v'^\alpha + B\dot{v}^\alpha + C\ddot{v}^\alpha$. Dynamic relevance of the last two features readily follow from the displacement optimization criterion, $\dot{v}_\alpha i^\alpha = 0$, and its derivative as $\ddot{v}_\alpha i^\alpha = 0$. Normalization of the coefficients $A$, $B$ and $C$ follows from the two considerations, namely, $f^\alpha v_\alpha = 0$, $f^\alpha i_\alpha = 0$, and also the consideration of constancy of proper mass $m_0$ in the dynamics. (Constant proper mass can be ensured from instant commoving inertial frame). Since, mass constancy corresponds to null value of $B$, the stated considerations lead to the same functionality of $f^\alpha$ as arrived through the nonlocal modification of LAD terms.

Now, it may be noted that the nonlocally modified term, $(2q^2/3c)\Theta i'^{\alpha}$, of the radiation reaction 4-force, besides having the space-like property ($i'^{\alpha}v_{\alpha}=0$), has the desirable feature of interacting with the displacement related LAD term in $f^{\alpha}$ as $[\ddot{v}^{\alpha}+\dot{v}^2v^{\alpha}]i'_{\alpha}=\ddot{v}^{\alpha}i'_{\alpha}=v_{\otimes}\dot{v}^2$. Finite interaction magnitude, $v_{\otimes}\dot{v}^2$, present in the radiative dynamics of accelerated charge is marker of nonlocally mitigated optimum displacement. Comprehensively, one thus finds that the 4-binormal component ($\dot{e}'^{\alpha}$) present in the LAD term interacts with $i'^{\alpha}$ as $i'^{\alpha}\sqrt{-\dot{v}^2}\dot{e}'_{\alpha}=v_{\otimes}\dot{v}^2$, where $\dot{e}'^{\alpha}\equiv(\dot{e}^{\alpha}-\sqrt{-\dot{v}^2}v^{\alpha})$ and $\ddot{v}^{\alpha}+\dot{v}^2v^{\alpha}=[\sqrt{-\dot{v}^2}\dot{e}'^{\alpha}-e^{\alpha}(\ddot{v}^{\beta}e_{\beta})]$; 4-jerk being expressible as $\ddot{v}^{\mu}\equiv d(\sqrt{-\dot{v}^2}e^{\mu})/d\tau=$ $\sqrt{-\dot{v}^2}\dot{e}^{\mu}-e^{\mu}(e^{\alpha}\ddot{v}_{\alpha})$ ($e^{\alpha}\equiv\dot{v}^{\alpha}/\sqrt{-\dot{v}^2}$ being 4-normal). In the nonstationary dynamics, the torsional characteristic implied in 4-jerk is thus mitigated by nonlocal field as $i'^{\alpha}\dot{e}'_{\alpha}=i^{\alpha}\dot{e}'_{\alpha}=-v_{\otimes}\sqrt{-\dot{v}^2}$. All other dynamically relevant local features, namely, 4-tangent $v^{\alpha}$, 4-normal $e^{\alpha}$ are not influenced in the dynamics as $i'^{\alpha}v_{\alpha}=0$, $e^{\alpha}i'_{\alpha}=0$. (4-tangent ($v^{\mu}$), 4-normal and 4-binormal with their mutual 4-orthogonalities as $\dot{e}'^{\alpha}e_{\alpha}=0$, $e^{\alpha}v_{\alpha}=0$, $\dot{e}'^{\alpha}v_{\alpha}=0$ are known to describe the local kinematic hypersurface). Now, noting that radiation 4-force is having the equivalent form of $f^{\alpha}\equiv(2q^2/3c)[\{\sqrt{-\dot{v}^2}\dot{e}'^{\alpha}-e^{\alpha}(\ddot{v}^{\beta}e_{\beta})\}+\Theta i'^{\alpha}]$, one writes electrodynamics as

$$m_0c\dot{v}_{\mu}-(q/c)F_{\mu\nu}v^{\nu}-(2q^2/3c)[\{\sqrt{-\dot{v}^2}\dot{e}'^{\alpha}-e^{\alpha}(\ddot{v}^{\beta}e_{\beta})\}+\Theta i'_{\mu}]=0 \qquad \text{(A2d)},$$

where, $\Theta=v_{\otimes}\dot{v}^2/(1+v_{\otimes}^2)$, $v_{\otimes}\equiv v_{\alpha}i^{\alpha}$, and $\dot{e}'^{\alpha}\equiv(\dot{e}^{\alpha}-\sqrt{-\dot{v}^2}v^{\alpha})$.

In the abridged local-nonlocal version of electrodynamics (eq.(A2d)), the dynamic variables are to be referred with their canonically averaged characteristics as these are realized with the weighted contribution of a family of coherently evolving paths $\{x_p^{\alpha}(\tau)\}$ of minimum displacement with optimum action in the dynamic course. In dynamics (A2d) a variable ($X$, say) is referred with its expectation value at an instant, which can be symbolically represented as $\langle X\rangle$. $\langle X\rangle$ is canonically averaged value of $X$ over the evolution volume at that instant. Expectation values of $f^{\alpha}$ is null when the dynamic evolution is coherently maintained by the ubiquitous nonlocal field. But, when the evolution decoheres at critical perturbation, $\langle f^{\alpha}\rangle$ acquires finite magnitude. It is important to mention here that in the coherently described dynamic course although the expectation value, $\langle\Theta i'_{\mu}\rangle$ on instant space-like surface is a null, $\Theta i'_{\mu}$ components in elemental 3-volumes of the surface need not be null to endorse existence of virtual exchange of radiation energy-momentum in field-particle interaction.

Now, since the Lorentz 4-force term in eq.(A2d), namely, $(q/c)F_{\mu\nu}v^{\nu}$ has null projection along 4-velocity ($v^{\mu}$) it is to be described in the kinematic hyperplane of 4-normal ($e^{\mu}$) and 4-binormal ($\dot{e}'^{\mu}$). Its canonically averaged projection on the 4-binormal, namely, $\langle F_{\mu\nu}v^{\nu}\dot{e}'^{\mu}\rangle$ can be written in an instant commoving inertial frame as $\langle \bar{E}\bullet\bar{\mathrm{e}}\rangle$. Since the periodically freely evolving binormal vector ($\bar{\mathrm{e}}$) makes null averaged value under a given external field $\bar{E}$, the quantity $\langle \bar{E}\bullet\bar{\mathrm{e}}\rangle$ is a null. In the averaged representation, Lorentz 4-force is therefore oriented solely along 4-normal ($e^{\mu}$). By considering this aspect and noting that $f^{\alpha}$ is made of two mutually 4-orthogonal components, namely, $-(2q^2/3c)(\ddot{v}^{\beta}e_{\beta})e^{\alpha}$ and $(2q^2/3c)[\{\sqrt{-\dot{v}^2}\dot{e}'^{\alpha}\}+\Theta i'^{\alpha}]$, the dynamics (A2d) can be expressed in the following two component equations:

$$\langle m_0 c\dot{v}_{\mu} + (2q^2/3c)(\ddot{v}^{\beta}e_{\beta})e_{\mu} - (q/c)F_{\mu\nu}v^{\nu}\rangle = 0 \qquad \text{(A3)},$$

and,

$$(2q^2/3c)\langle\sqrt{-\dot{v}^2}\dot{e}'_{\mu} + \Theta i'_{\mu}\rangle = 0 \qquad \text{(A4)}.$$

Eq.(A3) and (A4) respectively involve the direct and transverse components of the radiation 4-force $f^{\alpha}$. Their individual effects are narrated in the following subsections.

### 2.1.1 **Radiation 4-force involving 4-normal (direct components)**

Dynamic variables in the component equation (A3) are oriented along 4-normal. Eq.(A3) involves direct component of jerk 4-force, $(2q^2/3c)(\ddot{v}^{\beta}e_{\beta})e^{\mu}$. In instant commoving inertial frame, the equation is having the form, $\langle m_0\dot{\bar{\mathrm{v}}} - (2q^2/3c^3)(\ddot{\bar{\mathrm{v}}}\bullet\bar{\mathrm{e}})\bar{\mathrm{e}} - q\bar{E}\rangle = 0$, where $\ddot{\bar{\mathrm{v}}}\bullet\bar{\mathrm{e}} \equiv \ddot{\mathrm{v}}$. Jerk's presence results in impediment of the inertial force which is expressible as $\langle m_0\dot{\bar{\mathrm{v}}}[1\text{-}t_0 \mathrm{d}\ln(\dot{\mathrm{v}}/\dot{\mathrm{v}}_{\max})/\mathrm{dt}]\rangle$, where, $t_0 \equiv (2q^2/3m_0c^3) = c/\dot{\mathrm{v}}_{\max}$, $((2q^2/3m_0c^4)^{-1} = \dot{\mathrm{v}}_{\max})$. It implies that the instant proper frame observes impediment in momentum growth of the charged particle, which as discussed already is reflected in the growth rate of kinetic energy as ($\dot{\mathrm{k}}\text{-}t_0\ddot{\mathrm{k}}$). For the special case of uniformly accelerated dynamics ($\langle\ddot{\mathrm{v}}\rangle = 0$), the momentum impediment is absent.

### 2.1.2 **Radiation 4-force involving transverse components (4-binormal and nonlocally mediated term)**

Component equation (A4) represents dynamic balance of two transverse 4-forces respectively of torsion based 4-jerk component, $(2q^2/3c)\langle\sqrt{-\dot{v}^2}\dot{e}'^{\mu}\rangle$ and the nonlocally mitigating 4-force, $(2q^2/3c)\langle\Theta i'^{\mu}\rangle$ for optimum displacement in the dynamic course. (A4) expresses dynamic balance of nonlocally mitigated counteraction force to the torsion derived external perturbation. This description is evidently absent in the conventional formalism. As mentioned already, the torsional perturbation ingrained in jerk evolution will be the root cause of thermal relaxation in the nonstationary dynamics. In an instant commoving proper frame, the two involved 4-forces are having their respective components as $\dot{e}'^{\alpha}_{\text{proper}} \equiv [\dot{\text{e}}^0_{\text{proper}}, \dot{\bar{\text{e}}}_{\text{proper}}]/c$, $\dot{\bar{\text{e}}}_{\text{proper}} = \text{d}\bar{\text{e}}_{\text{proper}}/\text{dt}$, and $i'^{\alpha}_{\text{proper}} \equiv [\text{i}^0_{\text{proper}}, \bar{\text{i}}_{\text{proper}}]$; (scalar $-\dot{v}^2_{\text{proper}}$ is having the form, $-\dot{v}^2_{\text{proper}} = \dot{\text{v}}^2_{\text{proper}}/c^4$). In the frame, an observer will note the dynamic balance of torsion 3-force, $(2q^2/3c)\langle(\dot{\text{v}}/c^2)(\dot{\bar{\text{e}}}/c)\rangle$ and mitigating 3-force $(2q^2/3c)\Theta\bar{\text{i}}_{\text{proper}}$. The mitigating 3-force can be simplified further by replacing $v_{\otimes}$ (as $v_{\otimes} = \text{i}^0_{\text{proper}}$) in the quantity $\Theta$, $\Theta = v_{\otimes}(-\dot{\text{v}}^2/c^4)/(1+v^2_{\otimes})$. The replacement leads to $\Theta = (-\dot{\text{v}}^2/c^4)\text{i}^0_{\text{proper}}/[1+(\text{i}^0_{\text{proper}})^2] \equiv (-\dot{\text{v}}^2/c^4)(\text{i}^0/\text{i}^2)_{\text{proper}}$, $(\because i^{\alpha} i_{\alpha} \equiv (\text{i}^0)^2 - \text{i}^2 = -1)$. Thus, one expresses the 3-vector, $(2q^2/3c)\Theta\bar{\text{i}}_{\text{proper}}$ as $(2q^2/3c)(-\dot{\text{v}}^2/c^4)[(\bar{\text{i}}/\text{i})/(\text{i}/\text{i}^0)]_{\text{proper}}$, or, as $(2q^2/3c)(-\dot{\text{v}}^2/c^4)[\bar{\text{n}}_{\text{w}}/(\text{w}/c)]_{\text{proper}}$ where, $\bar{\text{w}}/c = \bar{\text{i}}/\text{i}^0$ is phase velocity ($\bar{\text{n}}_{\text{w}}$ being unit vector of $\bar{\text{w}}$, $\bar{\text{n}}_{\text{w}} = \bar{\text{i}}/\text{i} = \bar{\text{w}}/\text{w}$). The mitigating 3-force, $(2q^2/3c)\Theta\bar{\text{i}}_{\text{proper}}$, finally takes the local form as $(2q^2/3c^4)\langle -\dot{\text{v}}^2\rangle[\bar{\text{n}}_{\text{w}}/\text{w}]$. The dynamic force balance (eq.(A4)) in the local frame can be written as $(2q^2/3c^4)[\langle\dot{\text{v}}\dot{\bar{\text{e}}}\rangle - \langle\dot{\text{v}}^2\rangle(\bar{\text{n}}_{\text{w}}/\text{w})] = 0$, $(\infty \geq \text{w}/c \geq 1)$. This result shows that at criticality, where, $\lim_{w/c\to 1} \bar{\text{n}}_{\text{w}} = \bar{\text{n}}_{\text{rad}}$, ($\bar{\text{n}}_{\text{rad}}$ being radiation vector), the nonlocal counteraction against the jerk derived torsional perturbation attains highest magnitude. There, the nonlocal field compromises on the phase communication speed by attaining its lowest magnitude ($c$) as $\text{w}/c = 1$. Thus, against the torsion based 4-force, $(2q^2/3c^3)\langle\dot{\text{v}}\dot{\bar{\text{e}}}\rangle$, the counteracting 4-force is manifesting as recoil due to Larmor radiation as $-(2q^2/3c^4)\langle\dot{\text{v}}^2\rangle\bar{\text{n}}_{\text{rad}} \equiv -\dot{\bar{\Pi}}$. The critical balance leads to the equality, $\langle\dot{\bar{\text{e}}}\rangle_{cr} = \langle\dot{\text{v}}/c\rangle_{cr}\bar{\text{n}}_{\text{rad}}$; radiation recoil providing measure of jerk derived critical perturbation of the torsion type. The perturbation leads to stress yield under loss of coherence. The torsional relaxation out of stress yield in the nonstationary dynamics is an additional feature over the well known regular (Larmor) relaxation loss. Both losses are sustained in the nonstationary state at the expense of power from external supply. In this state, nonlocal field critically mitigates for optimum relaxation losses of the regular and disordered types.

Recoil momentum from regular relaxation is now recognizable as the signature of limiting nonlocal defense against perturbation; reactive defense from its virtual state below the criticality, finally shows off its existence while attaining the defense limit. Recoiled displacement has transverse component, the rate of which is proportional to its impact momentum to be expressed as $\bar{\text{v}}_{\text{trans}} \propto -\dot{\bar{\Pi}} t_0/m_0$. Since the recoil impedes on the jerk based torsion evolution of polarized vacuum field around the charged particle, the impediment involves a power term, which is expressible as $(2q^2/3c^3)\left|\dot{\mathrm{v}}\bar{\mathrm{e}}_{cr}\bullet\bar{\mathrm{v}}_{\mathrm{trans}}\right|$ $=\left|\bar{\dot{\Pi}}\bullet\bar{\mathrm{v}}_{\mathrm{trans}}\right|\equiv\dot{\mathrm{E}}_{\mathrm{HW}}$ (say). As mentioned already, heuristically argued dynamic balance of transverse force components in electrodynamics also points out this power expression. Considering the dynamically relevant radius ($\mathrm{r}_0 = ct_0 = 2q^2/3m_0c^2$) of the charge, the power density, $\dot{\mathrm{E}}_{\mathrm{HW}}/(4\pi\mathrm{r}_0^2)$ can thus be written as proportional to $(q^2/c\hbar)\left\langle\dot{\mathrm{v}}^2\right\rangle^2(\hbar/6\pi c^6)$. The exact nature of this proportional relation will be obtained by analyzing torsion driven damped oscillation of internal modes of the spherically polarized vacuum field of the accelerated charge (refer subsection 2.1.3). Even for the special case of uniformly accelerated linear motion of a charge ($\langle\ddot{\mathrm{v}}\rangle=0$), the torsion component $\left\langle\dot{\mathrm{v}}\dot{\bar{\mathrm{e}}}\right\rangle$ is finite indeed. With the uniform expectation value of acceleration, the case belongs to the category of nonstationary dynamics having relaxation losses of the ordered and disordered types.

It is to be noted that Larmor relaxation rate, $(2q^2\dot{\mathrm{v}}_{cr}^2/3c^4)\equiv\dot{\Pi}$, is $(2q^2\dot{\mathrm{v}}_{cr}/3m_0c^4)$ times the magnitude of inertial force ($m_0\bar{\dot{\mathrm{v}}}_{Cr}$). The dimensionless quantity $(2q^2\dot{\mathrm{v}}_{cr}/3m_0c^4)$ is a ratio of the two accelerations magnitudes, namely, $\dot{\mathrm{v}}_{cr}$ and $\dot{\mathrm{v}}_{\max}\equiv(2q^2/3m_0c^4)^{-1}=c/t_0$. This ratio suggests that recoil force would dominate over the inertia related dynamic property of the accelerating charge if acceleration exceeds the value of $\dot{\mathrm{v}}_{\max}$, which for an electron is about $4.8\mathrm{x}10^{31}\mathrm{ms}^{-2}$; for proton the $\dot{\mathrm{v}}_{\max}$ value is about 1836 times higher. Beyond $\dot{\mathrm{v}}_{\max}$, external field energy is however diverted significantly to the relaxation losses instead of manifesting as the increasing kinetic energy of the charge. Furthermore, one may note from the equality, $c\dot{\Pi}=(m_0c^2/t_0)(\dot{\mathrm{v}}/\dot{\mathrm{v}}_{\max})^2$, that at $\dot{\mathrm{v}}=\dot{\mathrm{v}}_{\max}$, Larmor relaxation occurs at the critically highest damping rate of $(m_0c^2/t_0)$. However, beyond the acceleration limit of $\dot{\mathrm{v}}>\dot{\mathrm{v}}_{\max}$, it is seen that the radiation power exceeds the critical feed rate of external power source.

***Nonlocal restraint on acceleration runaway in electrodynamics***

It is important to note that the presence of incessant nonlocal mitigation under critical loss of coherence restrains acceleration runaway as a possible solution of eq.(A3) in the particular situation of abrupt withdrawal of external field. In the stated situation, eq.(A3) apparently predicts the undesirable exponential growth of acceleration. Under abrupt withdrawal of the $\bar{E}$-field the local form of (A3), $\left\langle m_0\bar{\dot{\mathrm{v}}}-(2q^2/3c^3)(\bar{\ddot{\mathrm{v}}}\bullet\bar{\mathrm{e}})\bar{\mathrm{e}}-q\bar{E}\right\rangle=0$, being expressible as $\left\langle\dot{\mathrm{v}}-t_0\ddot{\mathrm{v}}\right\rangle=0$, ($t_0=2q^2/3m_0c^3$), allows exponential growth of acceleration as $\left\langle\dot{\mathrm{v}}\right\rangle_t=\left\langle\dot{\mathrm{v}}\right\rangle_{t=0}\mathrm{Exp}[\mathrm{t}/t_0]$. This as a possible solution of LAD-electrodynamics was in fact pointed out by Dirac himself [1] and also by other authors while giving possible resolutions as remedies of the evident crisis. In view of the presently explored nonlocal influence on radiation related force, it is prudent to examine whether the complementary equation (A4), that governs the nonlocal influence, supports or prohibits the exponential acceleration growth as a possible dynamic solution. As per Eq.(A4), the nonlocal field mitigates on the critically perturbing transverse component of the jerk force, and the mitigating effect is registered as radiation reaction as $\langle \bar{\dot{\mathrm{e}}} \rangle_{cr} = \langle \dot{\mathrm{v}}/c \rangle_{cr} \,\bar{\mathrm{n}}_{\mathrm{rad}}$. This result suggests that the acceleration value cannot exceed the limit given by $\langle \dot{\mathrm{v}} \rangle_{cr} = c\sqrt{\langle \bar{\dot{\mathrm{e}}} \rangle_{cr}^{2}}$, where $\bar{\dot{\mathrm{e}}} = \sum_{\mathrm{k},\omega_{\mathrm{k}}} \bar{\mathrm{e}}_{\mathrm{k}}(\text{-i}\omega_{\mathrm{k}})\exp(\text{-i}\omega_{\mathrm{k}}\mathrm{t})$. Representing the canonically averaged value of $\bar{\dot{\mathrm{e}}}$ as $\omega_0$, $\omega_0 = 2\pi/4t_0$, one finds that nonlocally mitigated acceleration cannot exceed the value of $c\omega_0$~7.54x10$^{31}$ms$^{-2}$. The ubiquitous presence of nonlocal mitigation on critically perturbed field-particle interaction thus rules out the exponential growth of acceleration as a possible dynamic solution under the abrupt withdrawal of external field. Further it is worthwhile to mention that with the abrupt withdrawal of external power source, the only possible source to sustain radiation relaxation of the accelerating charge is the evolution energy of the polarized charge sphere having the maximum value of the order of $m_0(c\omega_0 t_0)^2/2$ ~1x10$^{-13}$J. The limited evolution energy relaxes out rapidly after the withdrawal of external power and the charged particle then executes free motion ultimately.

### 2.1.3 Hawking-Unruh type thermal relaxation in accelerated dynamics

The radiation recoil from Larmor emission of the accelerated point charge leads to driven damped oscillation of the charge-polarized sphere of vacuum field having dynamically relevant radius of $r_0 \equiv (2q^2/3m_0c^2)$. As discussed at the beginning, the spherically polarized vacuum field of moving charge experiences the torque stress of $\bar{F} \equiv (2\mathrm{q}^2/3c^3)\dot{\mathrm{v}}\bar{\dot{\mathrm{e}}} = \bar{\Gamma}\dot{\mathrm{e}}/c$, where, $\bar{\Gamma} \equiv [r_o\bar{\mathrm{n}}_{\mathrm{r}} \times m_0\bar{\dot{\mathrm{v}}}] = r_o m_0 \dot{\mathrm{v}}\bar{\varepsilon}$, ($\bar{\varepsilon} \equiv \bar{\dot{\mathrm{e}}}/\dot{\mathrm{e}} = \bar{\mathrm{n}}_{\mathrm{r}} \times \bar{\mathrm{e}}$, $\dot{\mathrm{e}}$ being torsion rate and $\bar{\mathrm{n}}_{\mathrm{r}}$ being unit radial vector over the polarized sphere). In a recoil event, two of the three mutually orthogonal harmonic modes of spherical oscillator forming the canonically averaged plane are torsionally driven under the transverse field of electromagnetic reaction $-\dot{\bar{\Pi}}$; $\bar{\Gamma}\dot{\mathrm{e}}/c - \dot{\bar{\Pi}} = 0$. Using standard analytical results of driven damped oscillation [11] under critical perturbation one finds that energy relaxation occurs at the rate of $(2\varsigma m_0\omega_0)^{-1}\dot{\Pi}^2\zeta = \dot{\mathrm{E}}_{\mathrm{relax}}(\mathrm{say})$ . $\varsigma$, the damping parameter is unity for critically damped oscillation occurring with fundamental angular frequency $\omega_0$ of the harmonic modes. Multiplier, $\zeta$ represents recoil driven off-resonance factor; $\zeta = [(2\Delta\omega/\omega_0)^2 + 1]^{-1}$, where $\Delta\omega = \omega_0 - \omega_{\mathrm{rad}}$, $\omega_{\mathrm{rad}}$ representing frequency of recoiling photons in Larmor radiation. Noting that one full oscillation period involves sweeping over four times the radial dimension, one expresses $\omega_0$ as $\omega_0 = 2\pi/4t_0 = 2.5\times10^{23}$ s$^{-1}$, ($t_0 = 2q^2/3m_0c^3$ ~6.3x10$^{-24}$s). $\Delta\omega << \omega_0$ represents the near resonance cases, for which $\zeta \sim 1$, whereas $\Delta\omega \sim \omega_0$ represents off resonance cases for which $\zeta \sim 1/5$. $\dot{\mathrm{E}}_{\mathrm{relax}}$ has finite magnitude in nonstationary evolution having nonzero expectation value of acceleration. In fact, the above mentioned power loss expression in critical damping can be rewritten as $\dot{E}_{\mathrm{relax}} = [(\dot{\bar{\Pi}} \bullet \delta\bar{\Pi}/m_0)/\pi]\zeta$, where, $\dot{\bar{\Pi}} = (\dot{E}_{\mathrm{Larmor}}/c)\bar{\mathrm{n}}_{\mathrm{rad}}$ $\equiv (m_0c/t_0)(\dot{\mathrm{v}}/\dot{\mathrm{v}}_{\max})^2\bar{\mathrm{n}}_{\mathrm{rad}}$, and $\delta\bar{\Pi}/m_0 = \dot{\bar{\Pi}}t_0/m_0$ is the recoil velocity, ($\dot{E}_{\mathrm{Larmor}}$ being the rate of energy loss in Larmor radiation and $\dot{\mathrm{v}}_{\max} = c/t_0 \equiv 3m_0c^4/2q^2$). As mentioned already, $\dot{\mathrm{v}}_{\max}$ for electron is of the order of 10$^{31}$ms$^{-2}$. It is evident that as compared to Larmor radiation loss ($\dot{E}_{\mathrm{Larmor}}$) the damping power loss $\dot{E}_{\mathrm{relax}}$ is insignificant under normal acceleration values as can be seen by the ratio: $\dot{E}_{\mathrm{relax}}/\dot{E}_{\mathrm{Larmor}} = \zeta[(\delta\Pi/m_0c)/\pi] = \zeta\left(\dot{\mathrm{v}}^2/\dot{\mathrm{v}}_{\max}^2\right)/\pi$, $1 \geq \zeta \geq 1/5$. At the limiting acceleration, $\dot{\mathrm{v}}_{\max}$, polarized sphere is recoiled predominantly with radiations of near resonant frequencies, $\omega_0 \sim \dot{\mathrm{v}}_{\max}/c$ ($\omega_0 \approx 2.5\times10^{23}$ s$^{-1}$) attaining thereby the maximum value of the two power ratio as $\dot{E}_{\mathrm{relax}}/\dot{E}_{\mathrm{Larmor}} = 1/\pi$.

It can be seen that thermal power in the torsionally driven damping is expressible as $\dot{E}_{\mathrm{relax}} = \dot{E}_{\mathrm{Larmor}}\zeta\left(\dot{\mathrm{v}}^2/\dot{\mathrm{v}}_{\max}^2\right)/\pi \equiv \zeta[S_0\alpha\hbar\dot{\mathrm{v}}^4/6\pi^2c^6]$, where $\alpha \equiv q^2/c\hbar \sim 1/137$ is the fine structure constant, and $S_0 (= 4\pi c^2\tau_0^2)$ is the surface area of the dynamically significant sphere of radius, $c\tau_0$. $\dot{E}_{\mathrm{relax}}$ has the bi-quadratic functionality in acceleration indeed and it explicitly expresses the proportionally defined quantity $\dot{\mathrm{E}}_{\mathrm{HW}}$ which was discussed earlier. Considering that the time averaged disordered type relaxation rate $\dot{\mathrm{E}}_{\mathrm{HW}}$ occurs uniformly over the entire $4\pi$ solid angle of polarized sphere, the power density expression can be rearranged as ($\dot{E}_{relax}/S_0) = [\zeta(160\alpha)][\sigma(\hbar\dot{\mathrm{v}}/2\pi c\kappa_B)^4]$, where $\sigma = \pi^2\kappa_B^4/60\hbar^3c^2$, is the Stefan-Boltzmann constant. The power density corroborates to black body emission temperature of $\mathrm{T}_{\mathrm{rad}}(K) = (160\alpha\zeta)^{1/4}[\hbar\dot{\mathrm{v}}/2\pi c\kappa_{\mathrm{B}}]$, which can be approximated to $1.04\zeta^{1/4}[\hbar\dot{\mathrm{v}}/2\pi c\kappa_{\mathrm{B}}]$, where, $\kappa_{\mathrm{B}}$ is Boltzmann constant. $\zeta$-value assumes unity at the highest possible nonlocal defense, which is near the critical acceleration of $\dot{\mathrm{v}}_{\max}$. With lowering of acceleration, $\zeta$ reduces and converges to the lowest limit of 1/5, ($1/5 \leq \zeta \leq 1$). Nonetheless, $\mathrm{T}_{\mathrm{rad}}$, which is based on analysis from inertial frame of observation, has semblance to Hawking-Unruh radiation temperature [9,10]. As against the result of this analysis based on instant commoving inertial frame of accelerated charge, H-U radiation temperature is ascribed to warmness of surrounding vacuum field envisaged by accelerated observer's frame. For an electron, at the limiting acceleration ($\dot{v}_{max}$) of 4.8x10$^{31}$m.s$^{-2,}$, the thermal relaxation loss ($\dot{E}_{relax}$) as compared with Larmor radiation loss is about 22 percent. The relaxation would compare with the Larmor loss at an acceleration of 8.5x10$^{31}$ ms$^{-2}$ and temperature of 3.4x10$^{11}$K. These results may have significance in understanding the thermodynamic state of a cosmic system as it undergoes spontaneous symmetry breaking process with manifestation of ultrahigh energies.

It has been speculated by some authors that the H-U radiation temperature in thermal relaxation can be tested with gyrating electron in cylindrical penning trap [12]. Because of the achievable radial acceleration of 6x10$^{19}$g, (g is acceleration due to gravity on earth's surface), a temperature as high as 2.4 K is attainable in thermal relaxation from the gyrating charge inside the trap. The relaxation will put heat load inside the cryogenically cooled trap and warm up the cold surrounding. It has been envisaged that the warming effect should be reflected in line broadening of quantum jump spectroscopic study of spin states of gyrating electron [12].

**3. Quantum nature of nonlocally mediated canonical energy and momentum**

The canonical governance of nonlocal mediator $i^{\alpha}$ in coherent dynamics will be rationalized now by considering the supplementary condition (A2b), namely, $\pi_{\alpha}i^{\alpha} = \pi_{\alpha}w^{\alpha} = 0$. Recognizing that the velocity-like 4-vector, $w^{\beta} \equiv p_0[1, \overline{\mathrm{w}}/c]$, $1 \geq c/\mathrm{w} \geq 0$, corroborates to the phase communication property as was noted for free particle evolution, one generally expresses its spectral components as $w_k^{\beta} \equiv p_{0,\mathrm{k}}[1, (\omega/c\mathrm{k})\overline{\mathrm{n}}_{\mathrm{k}}]$. In the spectral space, $k^{\beta} \equiv [\omega/c, \overline{\mathrm{k}}]$, $(\overline{\mathrm{k}} = \mathrm{k}\overline{\mathrm{n}}_{\mathrm{k}}, \mathrm{n}_{\mathrm{k}}^2 = 1)$, one recalls now the perennial property of phase velocity as $k^{\beta}w_{\mathrm{k},\beta} \equiv 0$. A comparison of this property with the evolution, $\pi_k^{\beta}w_{k,\beta} = 0$, reveals the proportional connection of $\pi_{\mathrm{k}}^{\alpha}$ with the spectral 4-coordinates as $\pi_k^{\alpha} \propto k^{\alpha}$. Since this spectral connection of the nonlocal mediator with canonical 4-momentum stands irrespective of the particle, external field, and their interactions, the proportionality constant is versatile in nature. The constant having dimension of action can be ascribed with basic action unit, namely, the reduced Planck's constant, $\hbar$. The canonical 4-momentum thus evolves characteristically in the quantized form as $\pi_k^{\alpha} = \hbar k^{\alpha}$. The above discussion shows that the delocalized evolution in the dynamic course occurs with the quantized energy and momentum: $\pi^{\mu} = \sum_k c_k \pi^{\mu}(k) = \hbar \sum_k c_k k^{\mu}$. Field-particle interaction in the coherent evolution involves virtual exchange of energy-momentum in quantized forms. Conservation properties concerned in the interaction will be governed by canonical evolution characteristics to be discussed in the following section.

### 3.1 Eigen function of nonlocal evolution

Noting that in electrodynamic evolution, the quantized spectral 4-momentum $\pi_k^\alpha$ can be expressed as $\hbar k^\alpha = \pi_k^\alpha = m_0 c v_k^\alpha + qA^\alpha / c$, ($A^\alpha$ being 4-potential components of electromagnetic field), one represents 4-momentum evolution in scalar form as $\sum_k \sum_{k'} c_k c_{k'} [(\hbar k^\alpha - qA^\alpha / c)(\hbar k'_\alpha - qA_\alpha / c) - m_0^2 c^2 v_k^\alpha v_{\alpha,k'}] = 0$, with $c_k c_{k'} = \delta_{kk'}$, ($\delta_{kk'} = 0$ for $k \neq k'$, and $\delta_{kk'} = 1$ for $k = k'$); $v_k^\mu$ represents spectral velocity components. Noting that the summation $\sum_k \sum_{k'} c_k c_{k'} (v_k^\alpha v_{\alpha,k'})$ in the evolution equation represents the scalar $(v^\alpha v_\alpha)$, which is unity, the result corroborates to the dispersion property of the nonlocal mediator ($w^\alpha$) as $(\hbar k^\alpha - qA^\alpha / c)^2 = m_0^2 c^2$. The wave dispersion property governs the phase velocity ($(\omega/\bar{\mathrm{k}}) \equiv \bar{\mathrm{w}}_{\mathrm{k}}$) and group velocity ($\mathrm{d}\omega/\mathrm{d}\bar{\mathrm{k}}$) of matter waves associated with the evolution of scalar charged particle of mass $m_0$. The space-time evolution $\psi(x)$ is obtainable by Fourier transformation of dispersion relation. Under nonrelativistic limit, the evolution involves insignificantly small kinetic energy as compared to the proper energy ($m_0 c^2$). There, the stated dispersion relation can be simplified to $c\hbar \mathrm{k}_0 \text{-} q\phi \approx m_0 c^2 + [\hbar \bar{\mathrm{k}} \text{-} (q/c)\bar{\mathrm{A}}]^2 / 2m_0$. Coherent evolution of matter wave featuring spin can be described by considering the spectral velocity components $v_k^\mu$ as projections of the Dirac's bispinor matrices. $v_k^\mu$ is then expressible as $\bar{C}_{ka} \gamma_{ab}^\mu C_{kb}$, where, $C_{kb} = c_k \hat{u}_b$, $\bar{C}_{ka} = \hat{u}_c^\dagger c_k^\dagger \gamma_{ac}^0$; $\hat{u}_s$ ($s = 0,1,2,3$) are base components of the bi-spinor space. In general, the $\psi(x)$ evolutions are linear endorsing energy-momentum conservation with unique existence of the charged particle in the evolution volume at any instant. Energy-momentum of field-particle interaction is integrally conserved in the evolution volume.

It is necessary to mention here that the dispersion relation expressed above needs modification for the additional energy-momentum ($O_k^\mu$) involved in the virtual exchange of radiation involved in coherent evolution. As an added term to the canonical 4-momentum, $O^\mu$ remains 4-orthogonal to displacement so that there is no net energy-momentum contribution of the virtual exchange process in displacement $dx^\alpha$ of the coherent evolution. In the dynamic space, $O^\mu$ evolves in the hyperplane of 4-normal ($e^\alpha = \sum_k a_k e_k^\alpha$), and 4-binormal vector ($\dot{e}'^\alpha / \sqrt{-\dot{e}'^2} = \sum_k b_k \varepsilon_k^\alpha$, $\dot{e}'^\alpha = \dot{e}^\alpha - \sqrt{-\dot{v}^2} v^\alpha$). Thus, one writes $O_k^\mu = a_k e_k^\mu + b_k \varepsilon_k^\mu$, $O_k^2 = a_k^2 + b_k^2$, $\sum_k a_k a_k^* = 1$, $\sum_k b_k b_k^* = 1$, with $a_k a_{k'} = \delta_{kk'}$, $b_k b_{k'} = \delta_{kk'}$. The involved coefficients $a_k$ and $b_k$ can be normalized by considering the quantized canonical 4-momentum, $\hbar k^\mu = (m_0 c v_k^\mu + O_k^\mu) + (q/c)A^\mu$. The term $(m_0 c v_k^\mu + O_k^\mu)$ in the quantized 4-momentum having the magnitude of $\sqrt{m_0^2 c^2 + O_k^2}$ corroborates to an effective mass $M_k$, ($M_k = \sqrt{m_0^2 + O_k^2 / c^2}$); $M_k$ is involved in the wave dispersion feature of nonlocal mediation. Since the coherent evolution is described in instant proper frame, the components of 4-normal ($e^\alpha$) and 4-binormal vector are expressible in the local frame as $e^\alpha = [0, \overline{\mathrm{e}}]$, and $\varepsilon^\alpha = \dot{e}^\alpha / \sqrt{-\dot{e}^\beta \dot{e}_\beta}$, $\varepsilon^\alpha \equiv [0, \overline{\varepsilon}]$, where, $\overline{\varepsilon} \equiv \overline{\mathrm{n}}_\mathrm{r} \times \overline{\mathrm{e}}$, ($\overline{\mathrm{n}}_\mathrm{r}$ being unit radial vector of the polarized charge); $\mathrm{e}^2 = 1$, $\varepsilon^2 = 1$, and $\overline{\varepsilon} \bullet \overline{\mathrm{e}} = 0$.

The canonical rules in the delocalized description of dynamics as arrived by using classical framework with the added criterion of nonlocal mediation characteristics of rendering minimized relaxation energy in critically perturbed state do conform to the well established quantum mechanical prescriptions. It is worthwhile to mention here that the presented description of delocalized dynamics has similitude with the well known Feynman's approach of expressing quantum mechanical probability $|\psi(\overline{r})|^2$ in a dynamic passage for arrival of a coherently evolving particle at the local coordinates, $\overline{r}(x, y, z)$. According to the approach, the probability is expressed as summed over the phase histories of all possible paths meeting at the prescribed location simultaneously [13].

### 3.2 Local realisms under loss of coherence in dynamic evolution

In the coherently evolving space, the overall unit probability of existence of particle at an instant is realized through normalized $\psi(x)$ as $\int |\psi(x)|^2 d^3x = 1$. The coherent interaction of field and particle involves exchange of virtual quanta instantly across the evolution space volume maintaining energy-momentum conservation principle. The unitary evolution continues unless the mediating nonlocal field fails to guard the coherent state against critically high perturbation. Mediated virtual course of energy-momentum exchange endorsing the probabilistic existence instantly at all locations within a large evolution space is however cut short abruptly with failure of the coherent governance. Failure of coherent defense leads to marked change in the evolution 3-volume instantly with spontaneous radiation event; radiating boundary defining the changed evolution volume. There is an abrupt adjustment of the delocalized exchange activity with change in quantum state of the interacting field and particle and the change is reflected in energy-momentum loss/gain in the radiation event. The radiation event and consequent recoil effect on concerned particle are indeed signatures of the limiting defense reaction from mitigating nonlocal defense while yielding to the perturbation. The charged particle shows up its momentary presence with recoiled momentum at the site of wave vector collapse. Although nonlocal field promptly resumes its defense effort in guarding coherency of the changed state, the effort need not succeed under sustained criticality of external perturbation. Sustained criticality leads to nonstationary state of dynamics of the concerned particle with incessant radiation emission/absorption.

In repeated experimental runs of making wave collapse events under replicated coherent evolution histories in well defined perturbing field, the average occurrence frequency of the event in a given volume element of evolution space corroborates to the coherent probability $|\psi(\bar{r})|^2$. It implies that the nonlocally mediation accomplishes delocalized virtual exchange of energy-momentum coherently across the evolution volume at any instant maintaining the irrevocable probability rule. Instant phase communication speed keeps coherence in the virtual modes mediating the energy momentum to every location inside the volume. In coherently evolved dynamic course this virtual phase communication continues on successive space-like surfaces till reaching the event witnessing surface where the mediation abruptly fails to sustain the widespread coherence in phase communication across. The event making surface witnessed abrupt changeover of instant communication from the widespread to a limited evolution volume bounded by the radiating boundary marking the decoherence event. The attainment of critically high perturbation at the marked region of space-like surface led to instant changeover of evolution volume thereby depriving other regions of the coherent phase communication outside the radiation boundary. The instant changeover of the probability rule to a particular location of the event making surface needs to be reconciled with finite attainment of phase communication speed $c$ ($\mathrm{w} \rightarrow c$) at criticality. The compromised mediation speed though helped achieving maximum possible nonlocal defense against perturbation, distant locations outside the eventful radiation boundary are hardly informed of the moment of localized event. However, nature apparently has alternative means to make the event related instant phase communication to distant locations. This is unquestionably proven by noting the angular correlations simultaneously in nonlocally entangled spin paired states of two oppositely moving photons/particles in their physically well separated locations [14]. Nonlocally mediating ubiquitous vacuum field described in quantized space can involve the widespread geodesics network as parallel transmission channel for the instant phase communication before, during and after a wave collapse event in field-particle evolution. During quantum state measurement of one particle/photon of the entangled pair, the heavy probe could be hardly perturbed while receiving little relaxing stress of the decoherence event at its interacting tip which is coherently connected by the geodesics network to all other locations of the widespread space-like surface. The unperturbed and coherently entangled geodesics network instantly transmits the phase information of relaxing stress to all other space points irrespective of their distances. As the consequence, the distantly spin correlated member of the particle/photon pair could receive information about the local failure of nonlocal defense instantly and accordingly makes changeover from the mixed state of spin angular momentum following conservation rule integrally over the evolume volume.

## 4. Summary and conclusion

The ab-initio formulation of electrodynamics made through abridged analysis of local and nonlocal realisms has added versatility to the classical approach to address quantum evolution in perfectly coherent state and also, in the critically perturbed state in external field with loss of coherence. It formulates electrodynamics considering the additional requirement of minimizing self field loss as radiation from accelerated charged particle so that the particle completes its dynamic course with optimal use of power from applied field. The requirement could be met by delocalized description of dynamic course by involving a bunch of infinitesimally differed paths whose orchestrated evolution is nonlocally mediated in accomplishing minimum displacement with stationary action in the course. In the dynamic course, while the displacement criterion optimizes radiation energy loss (if any), the endorsement of stationary action leads to the dynamic description in delocalized version with a definite evolution characteristics corroborating to the quantum mechanical prescription. The classical description on a world line does not ensure meeting the two requirements indicated above. The coherent coordination among the minimally displacing paths as mediated by ubiquitous nonlocal field is shown to have unitary evolution characteristics that govern incessant exchange of quantized energy-momentum in the field-particle interaction. As for the ubiquitous nonlocal mediation characteristics, omnipresent vacuum field constituted of virtual oscillators with zero-point energies $\hbar\omega/2$ covering a long continuous frequency range of $0 \le \omega \le \pi c/\lambda\!\!\!^{-}$ provides the underlying support; $\lambda\!\!\!^{-}$ being basic dimension of quantized 4-space (refer Appendix-1). The dynamic observables are understood by canonical averaging of their delocalized involvements in the evolution. In the coherently interacting state, there is no net energy loss/gain in the form of radiation relaxation. Evolution characteristics are assigned there with the expectation values of energy, momentum, and angular momentum; canonically averaged values in the respective cases are time independent. Canonical rule of coherent evolution, that endorses null expectation values of acceleration and its derivatives, ensures the absence of the radiation related force. Nonstationary states in contrast manifest with the loss of coherence at critically high perturbation from external fields that includes fluctuating vacuum field. There, acceleration with jerk components assumes finite expectation values to result in radiation relaxations. The observable dynamics manifests because of limitation of the omnipresent nonlocal defense to protect the coherent existence. The defensive reaction to perturbation is limited by its inverse dependence on phase communication speed involved in coherent mediation. As perturbation develops, the incessantly reacting nonlocal field jacks up its defense by maneuvering on phase communication speed in the evolution. However, by attaining the lowest possible value of phase velocity the nonlocal reaction could defend coherent state critically up to an upper limit of the perturbation, which is shown to be derived from transverse components of jerk force at a given acceleration; perturbation imparting shear stress across spherically polarized vacuum interface of the moving charge. The criticality leads to decoherence with stress yield under the loss of dynamic balance between the two opposing forces; perturbation superseding defensive reaction from nonlocal field. As the result, both order (Larmor) and disorder (Hawking-Unruh) types of relaxations manifest. The disordered relaxation rate as compared to that of the ordered type is insignificantly small unless acceleration is exceptionally high. The relaxations mark the failed but sustained effort of nonlocal defense in the nonstationary state; disordered relaxation marks the spontaneity in decoherence. Recoil effect from Larmor relaxation providing measure of the critical perturbation, stands for the nonlocal defense limit. The normally prevalent virtual reaction of nonlocal defense surfaces out at critical perturbation with the relaxation phenomenon at signal speed; criticality signifies confluence of nonlocal and local activities of natural phenomena manifesting with the characteristic signal speed limit, $c$. Under sustained perturbation with stress yields, the field-particle system executes nonstationary evolution with the characteristic dynamic properties of finite acceleration, and jerk in the general cases. In transient perturbation, the system makes transition, involving spontaneous release of quantized radiation, to a coherently evolving new stable state. Besides revealing the dynamic features in coherent state and in critical state with loss of coherence, the ab-initio formulation could be successfully extended to rationalize the nonlocally mediated geodesic evolution characteristics in curved space. The evolution features of the nonlocally entangled geodesics could be obtained. Mediated geodesics network characteristically describes free motion of a test particle in the delocalized passage in between arbitrarily selected space-like surfaces having time-like separation as was considered in defining a dynamic course in external field. Some essential features of the geodesics characteristics are appended with this text [Appendix-1 ]. Various efficacies of the quantized space described by the geodesics network is detailed elsewhere [15].

Critical defense of nonlocal field governed by angular acceleration ($\dot{\mathrm{v}}^2/c^2$) was noted in the expression of regular (Larmor) relaxation loss in electrodynamics. Consideration of this helps defining the maximum possible value of acceleration which corroborates to the uppermost limit in defending coherent interaction of field-particle system. For a charged particle, the acceleration magnitude, $\dot{\mathrm{v}}_{\max} \equiv (2q^2/3m_0c^4)^{-1}$, decides the limit in electromagnetic interaction. Beyond this limit the relaxation phenomena affect the recovery process of lost self energy from the external field within the dynamically relevant time (~6.3x10$^{-24}$s). The existence of uppermost acceleration limit and the imminent relaxation loss in the nonstationary situation suggests that the reportedly encountered acceleration runaway problem in the conventional electrodynamics has little significance. Under abrupt withdrawal of external field, the jerk aided rising acceleration reduces to null value by critically mediated relaxation processes within the dynamically relevant time.

The ab-initio formulation, besides addressing acausality issue from acceleration runaway has led to several spinoffs following the revelation of quantized energy-momentum evolution property that rules delocalized dynamic course. While establishing rationale behind coherence and decoherence in evolution, it reveals the generality that energy relaxations perpetually associate with the defense limit of nonlocal field with loss of coherence. Relaxations of both ordered and disordered types respectively proportional to the angular acceleration ($\dot{v}^2/c^2$) and to the squared value $(\dot{v}^2/c^2)^2$ have been revealed. Hawking and Unruh radiations could be reasonably understood as offshoot of decoherence phenomenon, where disordered relaxation invariably manifests out of nonstationary evolution, be it in spontaneous manifestation of particle-antiparticle pairs from gravitationally perturbed vacuum field at event horizon of black hole, or, in warmed up surrounding as notable reality from accelerated state of an observer. (Finite acceleration of observer points to nonstationary state of his frame). For both the cases, this study has worked out a lab observer's experience of the thermal state in accelerated motion. In lab frame, nonstationary dynamic state generally showed typical decoherence characteristics, namely, the accelerated motion results in disordered relaxation loss with null chemical potential. The stated reality bears similitude with the accelerated observer's experience of his warm surrounding with blackbody radiation characteristics.

The relaxation features noted in electrodynamics could be generally extended to macro-world with a predictable ratio of power losses of the ordered and disordered types in critically perturbed state of dynamic systems in external field. The stated prediction is shown to be applicable uniquely in all scales of nonstationary dynamic evolutions. The power ratio obtained on the basis of mitigating nonlocal reaction to critical perturbation is used, as a common thread, to explain frictional contributions of phonons and electrons in normal and superconducting states of metals [16]. Furthermore, cosmic evolution features could be rationalized by considering the evolution to be in critically nonstationary state of the gravitationally active space. By ubiquitous mitigating reaction of coherent vacuum field, the space is incessantly driven to expand against the inward gravitational pull under critical loss of coherence. Using the basic characteristics of nonlocal mitigation at criticality and relaxations thereof in decoherence, the less understood features such as dark matter and dark energy could be accounted and reasonably quantified for their proportions. Briefly, this aspect is discussed in Appendix-1.

The classically rooted formulation of ubiquitous nonlocal interplay in dynamics, as presented here, shows versatility in rejuvenating the axiomatic approach of describing the coherent evolutions with their critical limits. Though the description is made primarily with electrodynamics, it is generally applicable to other dynamics.

## 5. Acknowledgement

This study was initiated during the tenure (2014-17) in Raja Ramanna Fellowship position after superannuation from service (2075-2013) in Bhabha Atomic Research Centre, Mumbai. Author thanks authorities of the host institute for their interest in basic studies.

7. Appendix 1

**Spinoffs from dynamical analysis -Space features on local-nonlocal realisms**

As discussed in the main text (Section 2), the nonlocally mediated geodesics network is describable with spectral dispersion property involving a basic dimension, $\lambdabar$. The derived dispersion relation endorses free motion of test particle and characterizes the vacuum field with quantized energy having density of $\rho_{vac}^{0} \equiv 2\pi c\hbar / \lambdabar^{4}$. Matter free flat space portrays field evolution in its pure state. From thermodynamic consideration the internal energy density of the field represented by $\rho_{vac}^{0}$ endorses its negative pressure characteristics to be expressed by the equation of state, $p_{vac}^{0} / \rho_{vac}^{0} = -1$. Noting this characteristics, reportedly observed speedy expansion of the gravitationally active cosmos has been speculated to be an outcome of the repulsive mitigation effect of vacuum field on the nonstationary space evolution. The repulsive pressure against gravitational inward pull of matters is presently considered to be due to 'dark energy' that manifests with the same equation of state as that of the field, namely, $p_{DE}^{0} / \rho_{DE}^{0} = -1$, where $\rho_{DE}^{0}$ and $p_{DE}^{0}$ respectively represent energy density and pressure of the maximum possible repulsive dark (unknown) manifest of the pure field in mitigating external perturbation. With the maximum mitigating posture of vacuum field, the dark energy density $\rho_{DE}^{0}$ can be treated as equivalent to the field energy density $\rho_{vac}^{0}$. One thus rewrites the negative pressure as $p_{DE}^{0} = -\rho_{DE}^{0} \equiv -\rho_{vac}^{0} = -2\pi c\hbar / \lambdabar^{4}$. Dark energy is constituted of energy quanta of the vacuum field ($2\pi c\hbar / \lambdabar$). Justification of dark energy as another facet of the vacuum field will be transparent in the presented analysis.

Now, the widely used Friedmann-Lemaitre-Robertson-Walker space metric for describing kinematics of energy-momentum stresses involved in gravitationally active space can be extended for analyzing the matter free state of (pure) vacuum field. Considering the general form of stress-energy equation in FLRW space as, $(\dot{a}/a)^2 = (8\pi G / 3c^2)(\rho_M + \rho_{DE}) - c^2/a^2$, where $a$ is the scale factor in the kinematic description and $\rho_{DE}$ is energy density of the dark manifest of vacuum field in presence of matter density $\rho_M$, one can represent the particular case of flat space with null $\rho_M$ and null space curvature as $(\dot{a}/a)^2 = (8\pi G / 3c^2)\rho_{DE}^{0}$; $G$ being gravitational constant (~6.674 x$10^{-11}$m$^3$kg$^{-1}$s$^{-2}$). The particular case involves the maximum possible value ($\rho_{DE}^{0}$) of the repulsive dark manifest of vacuum field, and does not involve the 3-curvature related last term ($-c^2/a^2$) of the general equation. The field mitigates on gravitational perturbation with the maximum possible repulsion when its coherent evolution is critically affected; mitigation is spontaneous and incessant until coherent state is reinstated. For the pure field, noting that the equation of momentum stress balance as $(\ddot{a}/a) = -(4\pi G / 3c^2)[\rho_{DE}^{0} + 3p_{DE}^{0}]$, one can find identity between energy and momentum stress equations. The identity follows from the two notable characteristics of field evolution, namely, (i) the equation of state as $p_{DE}^{0} / \rho_{DE}^{0} = -1$, and (ii) the equality of squared angular velocity with angular acceleration, $(\dot{a}/a)^2 \equiv c^2 / \lambdabar'^2 = (\ddot{a}/a)$. The first characteristics endorses the fact that the evolution involves dark energy with repulsive effect on gravitationally perturbed space, while the second one endorses field in its pure state where its geodesics network is described in matter free flat space and is therefore not stressed and distorted by inward pull of gravity. It is noteworthy that the uniqueness of energy-momentum evolution has similitude with that of photon field in vacuum. In gravitationally perturbed state the second characteristics starts deviating from the stated relation between angular velocity and angular acceleration; deviation standing as a measure of curvature stress developed in the affine space as will be realized later in the text.

Noting that the energy balance equation correlates observable kinematic properties of cosmic sphere, the quantity $(\dot{a}/a)^2$ is to be assigned with the value of squared angular velocity of spherically expanding 3-space to be represented as $(\dot{R}/R)^2$, $R$ being radial stretch of the observable spherical space of volume $4\pi R^3/3$. Further, considering that the space volume can be fully packed with $N$ quantized cubes of vacuum field, each of volume $\lambda\!\!\!^{-3}$, one finds that the radial stretch and the angular velocity are respectively expressed as $R=(3N/4\pi)^{1/3}\lambda\!\!\!^{-}$ and $(\dot{R}/R)=(\dot{R}/\lambda\!\!\!^{-})(4\pi/3N)^{1/3}$. In pure state of the field, $(\dot{R}/R)^2$ represents angular acceleration of its mitigating reaction to external perturbation. As reflected in the expression, acceleration varies inversely with the spherical surface of cosmic space. In the spherically represented space, the scale factor related quantity $(\dot{a}/a)^2$ can be assigned with the basic (angular) acceleration feature as $(c^2/\lambda\!\!\!^{-2})(4\pi/3)^{2/3}\equiv(\dot{a}/a)$. The stated finite magnitude of $(\dot{a}/a)^2$ characteristically corroborates to maximum possible nonlocal reaction as the field encounters critically high perturbation. Thus, the energy balance equation as applicable to the pure field described in the spherical space is having the form, $(8\pi G/3c^2)\rho_{DE}^0=c^2/\lambda\!\!\!^{-\prime 2}$, where, $\lambda\!\!\!^{-\prime}=(3/4\pi)^{1/3}\lambda\!\!\!^{-}$, ($\rho_{DE}^0=\rho_{vac}^0$).

Now, recalling the equalities, namely, $\rho_{vac}^0=\rho_{DE}^0=2\pi c\hbar/\lambda\!\!\!^{-4}$, and $\lambda\!\!\!^{-\prime}=(3/4\pi)^{1/3}\lambda\!\!\!^{-}$, and substituting them in the energy-momentum equation of vacuum field one gets the value of $\lambda\!\!\!^{-\prime}$ as $\lambda\!\!\!^{-\prime}=(3/4\pi)^{1/6}(4\pi G\hbar/c^3)^{1/2}$ and therefore, $\lambda\!\!\!^{-}$ as $\lambda\!\!\!^{-}=(4\pi/3)^{1/6}(4\pi G\hbar/c^3)^{1/2}$. In the literature, the constant factor $(4\pi G\hbar/c^3)^{1/2}$ ~5.72x10$^{-35}$ m is generally referred as Lorentz-Heaviside version of Planck's length). It is important to note that the basic dimension $\lambda\!\!\!^{-}$ is involved in the expression of the quantized energy, $2\pi c\hbar/\lambda\!\!\!^{-}$ (~2.74x10$^{9}$J) and of the energy density $\rho_{vac}^0$ (~7.13x10$^{111}$Jm$^{-3}$) of the field, whereas $\lambda\!\!\!^{-\prime}$ has relevance in expressing kinematic properties (angular velocity and acceleration) of the observable cosmic sphere. Pure field can attain angular acceleration of $(c/\lambda\!\!\!^{-\prime})^2$~ 4.43x10$^{85}$s$^{-2}$. Energy density and kinematic properties of the field are enormously large as compared to those of the gravitationally active cosmic space. As against the enormous field energy density, the critical density ascribable to the observably near flat space of the universe containing evenly distributed gravitating matter(s) and energy in cosmological scale, is insignificantly small, namely, $\rho_c=(8\pi G/3c^2)^{-1}H_0^2$ (~7.75x10$^{-10}$ Jm$^{-3}$). Observational data corroborate to angular velocity of the expanding space as $(\dot{a}/a)_{\text{obs}} \equiv H_0$ = 70±3 kms$^{-1}$Mpc$^{-1}$(~2.19x10$^{-18}$ s$^{-1}$); $H_0$ being the Hubble-velocity.

As will be elaborated below, (pure) vacuum field is the precursor of gravitationally structured space of the observable universe which is observably expanding with near homogeneous and isotropic distribution of matters in cosmic scale. Gravitating matters were derived out of the field energy immediately following inflationary expansion phase of the perturbed beginning (symbolic big-bang) of universe. In the abrupt expansion course the evolving field underwent metamorphic transformation partially and resulted in the gravitationally active space of observable normal matters along with unseen dark matter and energy components. In the gravitating space, the presence of dark components in significantly large proportions as compared to normal matters is reportedly arrived through observations of cosmic microwave background and studies of kinematics gravitating heavenly bodies like galaxies and clusters of galaxies. Field characteristics of quantized space described above helps providing a fairly good rational about dark matter and dark energy. Recalling the characteristics, pure vacuum field in matter free space can be endowed with two distinguishable features: (i) the basic space fabrics of nonlocally mediated geodesics, which as delocalized feature is stretch over space and in pure state of the field the geodesics evolves in lowest (ground) state with quantized energy of $2\pi c\hbar/\lambda\!\!\!^{-}$ and energy density of $2\pi c\hbar/\lambda\!\!\!^{-4} \equiv \rho^0_{vac}$, and (ii) the virtual lattice network formed out of the nonlocally entangled geodesics framework having energy density of $2\pi^2 c\hbar/\lambda\!\!\!^{-4} \equiv \rho^0_{lattice}$ (say); $\rho^0_{lattice}$ value follows from the consideration of density of states of orderly lattice network definable by ever present quantum oscillators with zero point energies ($\hbar\omega/2$) within the limits of $0 < \hbar\omega \leq 2\pi c\hbar/\lambda\!\!\!^{-}$. The ubiquitous precursor field endowed with the basic oscillators with all possible range of their zero point energies mediates nonlocally all coherently evolving states of the quantum world of matters. Coherent evolution changes into nonstationary state with minimum relaxation loss when the field mediates critically in mitigating high external perturbation.

Vacuum field endowed with the characteristic ratio of energy densities of its two distinguishable features as, $\rho^0_{lattice}/\rho^0_{vac} = \pi$, critically mitigates on gravitationally perturbed space in the nonstationary evolution. According to the well validated Lambda-CDM model, the gravitationally active space manifested out of inflationary space expansion with the symbolic big-bang beginning of the universe. Ubiquitous vacuum field with its incessant mitigation characteristics using the intrinsic dark pressure $\rho^0_{vac}$ reacted to initial singularity in space-time leading to the abrupt inflationary space expansion. During incessant mitigating reaction over the inflationary phase space, the quantized network of geodesics and virtual lattice networks of the evolving nonstationary state gave birth to the gravitationally active space by making partial transformation of the field features. In the metamorphic transformation involving symmetry breaking process, the virtual lattice network of the field's partially affected portion resulted in the manifestations of elementary particles as ordered quantum states of baryonic and leptonic species that form the normal (visible) matters. Geodesic network of the affected portion also underwent some modification. The two modified field features along with a comparatively larger background of untransformed field formed the gravitationally active space. Pure vacuum field, as the ubiquitous precursor continues governing the mitigated expansion of the formed space. With mitigation thrust, the field incessantly continues transfer of its virtual lattice and geodesic networks in their $\pi:1$ proportion to occupy the expanding space volume with the incompressible lattice network. Within the progressively developing vast background space, the gravitationally active regions of galaxies and galactic clusters made of normal matters exist in association with stressfully distorted geodesic network in their vicinities; distortion of network resulting from the intense local gravity field. The structured space of galaxies and clusters of galaxies was formed by segregation of normal matters in due course of the cosmic evolution. Today's observable cosmos of nearly homogeneous and isotropic distribution of galaxies and clusters of galaxies with their ever increasing interspaces represents the settlement of nearly 13.8 billion light years evolution history.

The locally stressed geodesics network can be represented by occupation of higher quantum states having energies $\hbar\omega_{vac}^{*}$ greater than that of pure vacuum field, that is, $\hbar\omega_{vac}^{*} > 2\pi c\hbar/\lambda$. The higher state possibly corresponds to exotic quantum field having similitude of bosonic field and achieving higher density of states under the stressful confinement in the local regions. The geodesics' distortion led to loss of homogeneity in the local background field of the gravitationally active regions. The local segregation of distorted geodesics leads to reformation of background density as $\rho_{vac}^{0} \to \rho_{vac}^{bckgr} + \langle\Delta\rho_{vac}\rangle$, where $\langle\Delta\rho_{vac}\rangle$ represents the space averaged locally accumulated stress in vicinity of galaxies and their clusters. The localized stresses remaining distributed over the entire gravitationally active regions of space an average value of energy density $\langle\Delta\rho_{vac}\rangle$ is considered. Because of disproportionately large empty space as compared to the volume occupation of galaxies and their clusters, the reformed background density $\rho_{vac}^{bckgr}$ is quite close to $\rho_{vac}^{0}$, and $\rho_{vac}^{bckgr}$ can be approximated to $\rho_{vac}^{0}$. Thermodynamically, $V\Delta\rho_{vac}$, ($V$ being local space volume) represents the distorted geodesics stress. This elastic type of stress causes increase in local pressure at constant volume. The invisible (dark) gravitating matter's density, $\rho_{DM}$ is accountable by $\langle\Delta\rho_{vac}\rangle$, that is, $\rho_{DM} = \langle\Delta\rho_{vac}\rangle$. The space is defined now with the total gravitating matter density ($\rho_M$) of $\rho_{DM} + \rho_{NM}$ ($\rho_{NM}$ is average density of normal matter). $\rho_M$, accounts a fraction of the critical density of $\rho_c$ that meets flatness criterion of the space; $\rho_c$ is negligibly small as compared to $\rho_{vac}^{0}$. The observable space is reportedly found to be flat, where the normal matters consisting of heavenly bodies, including plasma, gases and dusts, are considered to have homogeneous and isotropic distribution in the cosmic scale of space.

Following the partial transformation of lattice network the spontaneous mitigation of ubiquitous precursor field led to transfer of normal matters along with incompressible virtual lattice network under the differential pressure (density) of $(\rho_{vac}^{0} - \rho_{vac}^{bckgr}) = \langle\Delta\rho_{vac}\rangle, \langle\Delta\rho_{vac}\rangle > 0$. The transferred lattice network reveals its presence by volume occupation causing repulsive thrust against inward pull of gravity and resulting in the space expansion. The expansion is observable now by noting progressive increase of interspaces among cosmic objects. Dark energy as an attribute of the mitigating thrust can be accounted by considering the differential density, $\langle \Delta\rho_{lattice} \rangle = \rho^0_{lattice} - \rho'_{lattice}$, where $\rho'_{lattice}$ is lattice density in the gravitating space; $\rho'_{lattice} < \rho^0_{lattice}$. The other attributes are the visible matter of density $\rho_{NM}$ and dissipated radiation of density $\rho_{dissp}$. The differential density can be expressed as $\langle \Delta\rho_{lattice} \rangle = (\rho_{NM} + \rho_{DE} + \rho_{dissp})$, where $\rho_{DE}$ represents average density due to dark energy. The nonstationary evolution course involved dissipative radiation as byproduct of the transformation step. The overall mitigation effect of ubiquitous precursor field on gravitationally perturbed space is revealed by the consistent space expansion characteristics; mitigating parameters being the two differential densities, $\langle \Delta\rho_{lattice} \rangle$ and $\langle \Delta\rho_{vac} \rangle$, which are infinitesimally small compared to the precursor's pressure ( $\rho^0_{vac}$ ).

Noting that the precursor (pure) vacuum field with its geodesics network based virtual lattice is spatially homogeneous, the constant ratio of its two characteristic features, namely, $\rho^0_{lattice} / \rho^0_{vac} = \pi$, can be approximately applied to the transformed part of the network. Thus the two transformed features, $\langle \Delta\rho_{lattice} \rangle$ and $\langle \Delta\rho_{vac} \rangle$ are having their ratio as $\langle \Delta\rho_{lattice} \rangle \div \langle \Delta\rho_{vac} \rangle \approx \pi$. Now, by considering already arrived expressions of the numerator and denominator as $\langle \Delta\rho_{lattice} \rangle = (\rho_{NM} + \rho_{DE} + \rho_{dissp})$, and $\langle \Delta\rho_{vac} \rangle = \rho_{DM}$ respectively, one writes $(\rho_{NM} + \rho_{DE} + \rho_{dissp}) \div \rho_{DM} \approx \pi$. Considering that the energy density of dissipation is negligible in comparison with $(\rho_{NM} + \rho_{DE})$ one rewrites $(\rho_{NM} + \rho_{DE}) \div \rho_{DM} \approx \pi$. Noting that the average densities, $\rho_{NM}$, $\rho_{DM}$ and $\rho_{DE}$, are referred in the cosmic scale of homogeneous and isotropic universe, arrived ratio leads to $(\Omega_{NM} + \Omega_{DE}) / \Omega_{DM} \approx \pi$. This result as such corroborates to $\Omega_{DM} \approx 0.24$ and therefore to $(\Omega_{NM} + \Omega_{DE}) \approx 0.76$. The obtained results are, however, somewhat different from the reported values obtained by analyzing the observed CMB angular power spectrum with the help of Lambda-CDM model [P A R Ade et al 2016 Planck 2015 results A&A 594 A20; https://en.wikipedia.org › wiki ›Lambda-CDM model]. The modeled estimates are $\Omega_{DM} = 0.2589 \pm 0.0057$ and $\Omega_{DM} + \Omega_{NM} = 0.3089 \pm 0.0062$. As compared to the reported $\Omega_{DM}$ of about 0.26, the presented low value of $\Omega_{DM} \approx 0.24$ of this study is possibly because of ignoring the internal energy transfer between the two characteristic energy states (geodesics and their network of virtual lattice) of vacuum field during the mitigated transformation process. Use of overvalued energy density ratio of the two states resulted in low $\Omega_{DM}$. The individual values of dark energy and normal matter contents endorsed by the combined result, $(\Omega_{NM} + \Omega_{DE}) \approx 0.76$, can be settled by considering thermodynamic analysis of the mitigated spontaneous transfer of quantized states of geodesics and their virtual lattice network to the gravitationally active space.

The transfer process can be modeled as flow of enthalpy from ubiquitous precursor field (that is, pure vacuum field) of high energy density to the gravitating space of relatively lower density ($\rho_{vac}^{bckgr}$). The flow process is adiabatically treated in terms of the pressure-volume ($p$-$V$) effects on the internal energy change of the precursor field's features, namely, quantized geodesics network and associated virtual lattice. The delocalized state of geodesic network governs the flow under the differential energy density (pressure) of $\rho_{vac}^{0} - \rho_{vac}^{bckgr} = \Delta\rho_{vac} > 0$. Following the equation of state of vacuum field, positive value of $\Delta\rho_{vac}$ corresponds to net repulsive pressure effect against the gravitational inward pull of matters; repulsive pressure being transmitted through the incompressible virtual lattice network of the field introducing the kinematics of space expansion. The expansion rate of cosmic sphere at an instant can be written as $V(t)_{\delta t \to 0}^{\lim}[\{(1+H_t\delta t+...)^3-1\}/\delta t] \approx 3V(t)H_t$, where $V(t)$ and $H_t$ respectively represent the volume of cosmic sphere, and the Hubble's expansion parameter at an epochal time $t$, $0 \geq t \geq -\tau_{age}$; $\tau_{\rm age}$ ~13.8 billion light years, is the age of universe and $t=0$ corresponds to this epoch where the expansion parameter is $H_0$, the Hubble's constant. When this expansion takes place under a differential pressure of $p(t) \sim \Delta\rho_{vac}$, the cumulative pressure-volume type of work involved in the whole space evolution history is expressible as $E_{\rm w} = 3\int_{t=\tau_{\rm age}}^{t=0} p(t)V(t)H_t dt$. Besides manifestation of the displacement work, there can be accumulation of elastic energy over the period as given by $E_{\rm elast} = \int_{t=\tau_{\rm age}}^{t=0} V(t)\dot{p}(t)dt$, $\dot{p} = dp/dt$. The $p-V$ effects in the spontaneous flow of field energy thus leads to the stress manifestations with and without work signatures; in elemental transfer process these are respectively represented by $p\delta V$ and $V\delta p$. These two signatures of vacuum field mitigation in the gravitationally active space are respectively derived from change in energy states of the two distinguishable field features, namely, the orderly virtual lattice network and delocalized geodesics framework. Energy change in the respective features are orderly space occupation with incompressible lattice structure by performing displacement work, $E_{\rm w}$, and localized distortion of geodesics structure in vicinities of massive heavenly bodies by registering accumulations of elastic stress, $E_{\rm elast}$.

Mitigating vacuum field with its very high internal pressure (~$\rho_{vac}^{0}$) pushes the incompressible quantized cell volumes ($\lambda\!\!\!^-{}^3$) one after another into the gravitating space and thereby accomplishes the volume displacement work. The quantized volume units $\lambda\!\!\!^-{}^3$ being transferred in synchrony with the space expansion kinematics, the pressure thrust on a unit cell proportional to $p(t)\lambda\!\!\!^-{}^2$ generates the Newtonian force of $[(\rho_{vac}^{0}/c^2)\lambda\!\!\!^-{}^3]\ddot{a}(t)$; $a$ being the scale factor. $(\ddot{a}/a)$ corroborates to the observable acceleration $(\ddot{R}/R)$ of the cosmic sphere of radius $R$. Thus, one writes, $p \propto [\rho_{vac}^{0}(\lambda\!\!\!^-/c^2)a](\ddot{a}/a)$, that is, $p(t) = k_0(\ddot{a}/a)$, where $k_0$ involves $\rho_{vac}^{0}\lambda\!\!\!^- a/c^2$ and a constant multiplier arising from the proportional representation. Now, over the epochal expansion period, acceleration need not be steady. When one considers time evolved acceleration kinematics, the jerk effect manifests as $(\partial p / \partial t)_V \equiv \dot{p} = k_0(\dddot{a} / a)$; $(\dddot{a} / a) \approx (\dddot{R} / R)$. Thus, parallel to the work performed by volume displacement in the homogeneous and isotropic space expansion across the cosmic volume, there is pressure build up throughout the volume by jerk's presence. In the cosmic volume, $V(t)$, the pressure evolution can be represented with $p(t) \approx k_0(\ddot{R} / R) = k_0\varepsilon' H_t^2$ and $\partial p / \partial t \approx k_0(\dddot{R} / R) = k_0\varepsilon' H_t^3$, wherein the acceleration and jerk in the gravitationally active space are interrelated with the angular velocity $(\dot{R} / R)$ as $(\ddot{R} / R) = \varepsilon'(\dot{R} / R)^2$, and $(\dddot{R} / R) = \varepsilon'(\dot{R} / R)^3$. The parameter $\varepsilon'$ represents deviation from its ideal value of $\varepsilon' = 1$; ideality pertains to the case of vacuum field in its matter free pure state ($\Omega_{DE} = 1$ and $\Omega_M \equiv \Omega_{DM} + \Omega_{NM} = 0$). Gravitating space involves fractional $\varepsilon'$ ($\varepsilon' < 1$) to represent lowering of angular acceleration in space expansion from the ideal value of $H_t^2$. (In this epoch of near flat space with very low density of gravitating matters, the magnitude of $\varepsilon'$ can be correlated with the matter content by using energy-momentum stress balance in FLRW metric as $\Omega_M = 2(1-\varepsilon')/3$; for $\Omega_M = 0.3089$, $\varepsilon' = 0.54$).

In the critically mitigated space expansion, the two cumulative energy signatures thus manifest with a ratio of $E_{\mathrm{w}} / E_{\mathrm{elast}} = 3 \int_{t=\tau_{\mathrm{age}}}^{t=0} p(t)V(t)H_t dt \div \int_{t=\tau_{\mathrm{age}}}^{t=0} V(t)\dot{p}(t)dt$. Substituting the values of pressure, $p(t) \approx k_0\varepsilon' H_t^2$, and its time derivative, $\dot{p} \approx k_0\varepsilon' H_t^3$, the stress ratio is expressible as $E_{\mathrm{w}} / E_{\mathrm{elast}} \approx 3 \int_{t=\tau_{\mathrm{age}}}^{t=0} k_0\varepsilon' H_t^3 V(t)dt \div \int_{t=\tau_{\mathrm{age}}}^{t=0} k_0\varepsilon' H_t^3 V(t)dt = 3$. Considering that the two distinguishable expansion energies, $E_{\mathrm{W}}$ and $E_{\mathrm{elast}}$ are respectively the manifested effects of the invisible forms of energy and matter, one rewrites the two energy ratio as $E_{\mathrm{w}} / E_{\mathrm{elast}} \equiv E_{DE} / E_{DM} \approx 3$; $E_{DE}$ and $E_{DM}$ being the repulsive dark energy and gravitating dark matter respectively. while $E_{DE}$ is endorsing the displacement work, $E_{DM}$ endorses the accumulated elastic stress. This result is equivalent to $\Omega_{DE} / \Omega_{DM} \approx 3$, and it can be compared now with the previously obtained result according to mitigation characteristics of the precursor field as $(\Omega_{NM} + \Omega_{DE}) / \Omega_{DM} \approx \pi$. Comparison shows that $\Omega_{NM} / \Omega_{DM} \approx \pi - 3$ and this leads to $\Omega_{DM} \approx 0.24$, $\Omega_{DE} \approx 0.72$, and $\Omega_{NM} \approx 0.04$. The presented abinitio formalism thus shows versatility in revealing the rationale behind invisible matter and energy features in the space evolution.